\documentclass[a4paper,11pt]{article}
\pdfoutput=1 

\usepackage{jcappub} 
\usepackage{xcolor,bm}
\newcommand{\sigmanu}{$\Sigma m_{\nu}$}
\newcommand{\summnu}{\Sigma m_{\nu}}
\newcommand{\thetal}{\bm{\tilde{\theta}}}

\newcommand{\lyaf}{Ly$\alpha$ forest}
\newcommand{\pd}{$P_\mathrm{1D}$}
\newcommand{\pdv}{$\tilde{P}_\mathrm{1D}(q_\parallel)$}

\title{\boldmath An emulator for the Lyman-$\alpha$ forest in beyond-$\Lambda$CDM cosmologies}

\author[a,1]{Christian Pedersen,\note{Corresponding author}}
\author[b,a]{Andreu Font-Ribera,}
\author[c,d]{Keir K. Rogers,}
\author[e]{Patrick McDonald,}
\author[a,c]{Hiranya V. Peiris,}
\author[a]{Andrew Pontzen,}
\author[f]{and An\v{z}e Slosar}

\affiliation[a]{Department of Physics and Astronomy, University College London, Gower Street, London, United Kingdom}
\affiliation[b]{Institut de F\'{i}sica d’Altes Energies (IFAE), The Barcelona Institute of Science and Technology, 08193 Bellaterra (Barcelona), Spain}
\affiliation[c]{Oskar Klein Centre for Cosmoparticle Physics, Department of Physics, Stockholm University,
AlbaNova University Center, Stockholm 10691, Sweden}
\affiliation[d]{Dunlap Institute for Astronomy \& Astrophysics, University of Toronto, 50 St. George Street, Toronto, ON M5S 3H4, Canada}
\affiliation[e]{Lawrence Berkeley National Laboratory, One Cyclotron Road,
Berkeley, CA 94720, USA}
\affiliation[f]{Physics Department, Brookhaven National Laboratory, Upton, NY 11973, USA}

\emailAdd{christian.pedersen.17@ucl.ac.uk}
\emailAdd{afont@ifae.es}
\emailAdd{keir.rogers@utoronto.ca}
\emailAdd{pvmcdonald@lbl.gov}
\emailAdd{h.peiris@ucl.ac.uk}
\emailAdd{a.pontzen@ucl.ac.uk}
\emailAdd{anze@bnl.gov}

\abstract{
Interpreting observations of the Lyman-$\alpha$ forest flux power spectrum
requires interpolation between a small number of expensive simulations. 
We present a Gaussian process emulator modelling the 1D flux power spectrum as
a function of the amplitude and slope of the small-scale linear matter power
spectrum, and the state of the intergalactic medium at the epoch of interest
($2 < z < 4$).
This parameterisation enables the prediction of the flux power spectrum in
extended cosmological models that are not explicitly included in the training set,
eliminating the need to construct bespoke emulators for a number of extensions
to $\Lambda$CDM.
Our emulator is appropriate for cosmologies in which the linear matter power
spectrum is described to percent level accuracy by just an amplitude and
slope across the epoch of interest, and in the regime probed by eBOSS/DESI data.
We demonstrate this for massive neutrino cosmologies, where the emulator is able
to predict the flux power spectrum in a $\Sigma m_\nu=0.3$ eV neutrino cosmology
to sub-percent accuracy, without including massive neutrinos in the training simulations.
Further parameters would be required to describe models with sharp features in the
linear power, such as warm or light axion dark matter.
This work will facilitate the combination of upcoming DESI data
with observations of the
cosmic microwave background, to obtain constraints on neutrino mass and
other extensions to $\Lambda$CDM cosmology.
}

\begin{document}
\maketitle

\flushbottom

\section{Introduction}
\label{sec:int}

The Lyman-$\alpha$ forest (\lyaf) is a series of absorption features in the
spectra of high-redshift quasars, which is caused by the presence of intervening
neutral hydrogen.
During the last decade, the Baryon Oscillation Spectroscopic Survey
(BOSS, \cite{BOSS2013}) and its extension eBOSS \cite{eBOSS2016} have obtained
over 200,000 quasar spectra.
These surveys were able to measure the 3D correlation of absorption
features and detect baryon acoustic oscillations (BAO) around $z=2.3$
\cite{eBOSS2020}.
In addition, the same dataset has been used
to provide precise measurements of the flux power spectrum along the line
of sight, \pd\ \cite{Nathalie2013,Chabanier2019}.
In this paper we present a framework to interpret these \pd\ measurements
and obtain robust constraints on cosmological parameters.
Starting in 2021, the Dark Energy Spectroscopic Instrument
(DESI, \cite{DESI2016}) will observe 700,000 \lyaf\ quasar spectra,
and will provide the most precise measurement of the \pd\ to date.
\pd\ measurements are a unique probe of the clustering
of matter on megaparsec scales
\cite{Croft1998,Croft1999,McDonald2000,Gnedin2002,Croft2002,Viel2004a,McDonald2005}.
In particular, measurements from BOSS, eBOSS and DESI\footnote{\pd\ measurements from high-resolution spectra of $z>4$ quasars
have been used to study the reionisation epoch
\cite{Onorbe2017b,Walther2018,Boera2019}
and to constrain dark matter models
\cite{Viel2013,Irsic2017b,Irsic2017c,Rogers2020a,Rogers2020b}.
We do not discuss these measurements here.}
are sensitive to the clustering of matter on scales of
$0.2\,\mathrm{Mpc}^{-1} < k < 3\,\mathrm{Mpc}^{-1}$ \cite{Chabanier2019b}.
This information is highly complementary to other probes, such as observations
of the cosmic microwave background (CMB), and combined analyses have provided
some of the tightest constraints on the shape of the primordial power spectrum
and upper limits on the neutrino mass scale
\cite{Phillips2001,Verde2003,Spergel2003,Viel2004b,Seljak2005,Seljak2006,
Bird2011,Nathalie2015,Nathalie2015b,PD2020}.

Obtaining cosmological constraints from \pd\ measurements requires
theoretical modelling.
Perturbative approaches are not sufficiently accurate on the small scales to which
the \pd\ is sensitive.
Moreover, the absorption features are also sensitive to the state of the
intergalactic medium (IGM), which is dependent on baryonic physics and
the ultra-violet background \cite{McQuinn2016}.
Modelling these effects simultaneously requires expensive
hydrodynamical simulations, each costing $\sim 10^5$ CPU hours,
which means, in practice, that only a small number ($\lesssim 100$) of simulations
can be run for a given analysis. This is far fewer than the
$\sim 10^6$ likelihood evaluations necessary for common sampling
techniques, e.g., Markov chain Monte Carlo methods (MCMC),
to provide robust statistical constraints on parameters.
A solution to this problem has been to construct frameworks
that interpolate between simulations.
Similar problems exist in other
areas of cosmology, such as in analysis of galaxy surveys, where
Gaussian processes have been used to interpolate probabilistically
(or \textit{emulate}) the galaxy power spectrum \cite{Kwan2015}
and correlation function \cite{Zhai2019} as functions of
cosmology from a small number of expensive simulations.
Analysis of the \pd\ over the last decade most commonly used
quadratic polynomial interpolation \cite{Viel2006,Borde2014};
however recent work has demonstrated
the advantages of using a Gaussian process emulator to perform the
interpolation \cite{Walther2018,Murgia2018,Bird2019,Rogers2019,Takhtaganov2019,Rogers2020a,Rogers2020b}.

Historically, approaches to combining small-scale clustering
information from the \lyaf\ with CMB observations fall into two categories.
Early work on the \lyaf\ recognised that the \lyaf\ probes the universe at an epoch
that is close to an Einstein de-Sitter (EdS) model,
and therefore considered the cosmological information in the \lyaf\ to be concentrated in
the linear matter power spectrum \cite{Croft1998,McDonald2000,Croft2002,Gnedin2002}.
The procedure was then to constrain the $z\sim3$,
linear matter power spectrum on megaparsec scales from \lyaf\ observations
\cite{Croft1999,McDonald2000,Viel2004a,McDonald2005},
and combine these measurements with the CMB to constrain cosmological parameters
\cite{Phillips2001,Verde2003,Spergel2003,Viel2004b,Seljak2005,Seljak2006}.
More recent analyses \cite{Viel2006,Viel2010,Borde2014,Nathalie2015,Nathalie2015b,PD2020}
have instead modelled the \pd\
as a function of $\Lambda$CDM parameters, plus neutrino mass in the case of
studies into massive neutrinos.

The work we present here returns to the original approach. 
The motivation for this is twofold.
First, the effects of $\Lambda$CDM parameters on the
\lyaf\ are strongly degenerate with one another.
Minimising parameter degeneracies is important at all
stages of the analysis.
Degeneracies can be artificially broken by interpolation
errors, and lower dimensional parameter spaces allow for
a more efficient distribution of simulations.
Second, there are
several well-motivated extensions to $\Lambda$CDM that will
further characterise physical properties of the universe,
such as the number and mass scale of the neutrinos, or running
of the spectral index (see section 7 of ref. \cite{Planck2018}
for further examples).
When combined with CMB observations, the small-scale
clustering information provided by the \pd\ provides constraints
on these and many other extensions to $\Lambda$CDM.
However, given the large number of prospective parameter
spaces that one would want to explore, it is
impractical to construct an emulator for each.
A potential solution is to emulate the \pd\
in a parameter space that can simultaneously describe the effect of these
extended models on the flux power spectrum.

In this work we investigate this idea by constructing
a Gaussian process emulator modelling the \pd\ with just two
parameters in the cosmological sector: the amplitude
and slope of the linear matter power spectrum at the epoch of interest.
We demonstrate that this emulator can accurately predict the \pd\ in
massive neutrino cosmologies without including massive neutrinos
in the training simulations.

As we will show in section \ref{sec:param}, parameterising the linear
power spectrum in terms of just an amplitude and slope well describes
two commonly investigated extensions to $\Lambda$CDM, neutrino
mass \cite{Julien2006,Julien2014} and curved cosmologies \cite{Planck2018},
over the relevant range of length scales.
This is because these cosmologies induce smooth variations in the
small-scale linear power spectrum.
Many proposed alternatives to cold dark matter
(such as light axion \cite{Hu2000,Hui2017}, self-interacting \cite{Vogelsberger2016}
or sterile neutrino \cite{Koenig2016,Adhikari2017} dark matter)
feature a sharp suppression of the linear matter power spectrum
on small scales, which this parameterisation does not describe.
High resolution \pd\ measurements \cite{Irsic2017,Boera2019} have
been used to put bounds on properties of these models
by constraining this cutoff scale \cite{Viel2013,Irsic2017b,Irsic2017c,Murgia2018,PD2020,Rogers2020b}.
Recent results bound the cutoff scale at wavenumbers higher
than those probed by the \pd\ measured from BOSS/eBOSS/DESI,
which is the focus of this paper.
Therefore, while in principle our parameterisation of the linear power spectrum can
be extended to include a cutoff scale, we leave this investigation for future work.
    
The outline of this paper is as follows: in section
\ref{sec:param}, we describe the parameterisation
of the emulator in terms of the linear matter
power spectrum and state of the IGM. Section
\ref{sec:sims} describes the suite of simulations,
and the post-processing steps.
In section \ref{sec:emu}, we detail the construction
of the Gaussian process emulator and validate the predictions on
test data.
In section \ref{sec:sampler}, we run MCMC chains
to obtain constraints on the primordial power
spectrum from simulated mock data.
We discuss our results in section
\ref{sec:con}.

\section{Parameterisation}
\label{sec:param}
In this section we describe the parameterisation of our emulator.
The absorption features that constitute the \lyaf\ are caused
by the presence of intervening neutral hydrogen along the lines
of sight to quasars. The density of this neutral hydrogen
is determined
both by fluctuations in the baryon density which trace the
underlying dark matter field, and the thermal and ionisation
state of the IGM. We use a total of six parameters to describe
these effects: two to describe the clustering of matter and four
to characterise uncertainty in the state of the IGM. These are discussed in sections
\ref{ss:linpower} and \ref{ss:igm} respectively.

\subsection{Linear matter power spectrum}
\label{ss:linpower}

The \lyaf\ is sensitive to the matter density field.
We parameterise the \pd\ with the amplitude \(\Delta^2_p(z)\) and slope \(n_p(z)\) of the combined
cold dark matter + baryon linear power spectrum\footnote{Note that
in the case of cosmologies with massive neutrinos, we do not
include the clustering of the neutrinos in this quantity.}, 
$P_\mathrm{lin}$, defined as:
\begin{equation}
    \Delta^2_p(z)=k^3P_\mathrm{lin}(k,z)|_{k=k_p},
    \label{eq:Delta2_p}
\end{equation}
\begin{equation}
    n_p(z)=\mathrm{d}\mathrm{ln}P_\mathrm{lin}(k,z)/\mathrm{d}\mathrm{ln}k |_{k=k_p},
    \label{eq:n_p}
\end{equation}
where we use a pivot scale $k_p=0.7\: \mathrm{Mpc}^{-1}$
which was used in the measurement of the small-scale linear
matter power spectrum of ref. \cite{McDonald2005}.
To numerically evaluate $\Delta^2_p$ and $n_p$, we use
\texttt{CAMB} to generate a linear matter power spectrum
and fit a second order log polynomial over the range
$0.5 k_p < k < 2 k_p$.
We do not include parameters to describe changes
to the growth rate or expansion history.
This choice is motivated by the fact that in the redshift range
that gives rise to the \lyaf\ ($5>z>2$), the universe
is very close to Einstein-de Sitter (EdS) with $\Omega_m(z=2)\approx0.91$ and
$\Omega_m(z=5)\approx0.99$ for typical cosmological models allowed by the CMB.
In an EdS universe, the expansion rate evolves
$H(z)=\dot{a}/a\propto (1+z)^{3/2}$,
and the growth factor evolves proportional to $a$.
In particular, for the expansion history, there is a $1.2\%$ difference in
$H(z=2)$ between flat cosmologies with $h=0.67$ and $h=0.74$
at fixed physical density $\omega_m=\Omega_m h^2$,
which is the range of the current
tension between early and late universe constraints on $H_0$ \cite{Riess2016}.
This difference drops to $0.5\%$
at $z=4$ as we move deeper into the EdS regime.
Similarly, when we compare the logarithmic growth rates for cosmologies
with $\Omega_m=0.25$ and $\Omega_m=0.35$, we see a $2.1\%$ difference at
$z=2$ and a $0.4\%$ difference at $z=4$. We construct our
emulator under the assumption that the relationship between the matter
power spectrum and the \pd\ is insensitive to these slight changes
in expansion history and growth rate, and we consider the results shown
in sections \ref{sec:emu} and \ref{sec:sampler} a test of this assumption.

\begin{figure}
    \centering
    \includegraphics[scale=0.42]{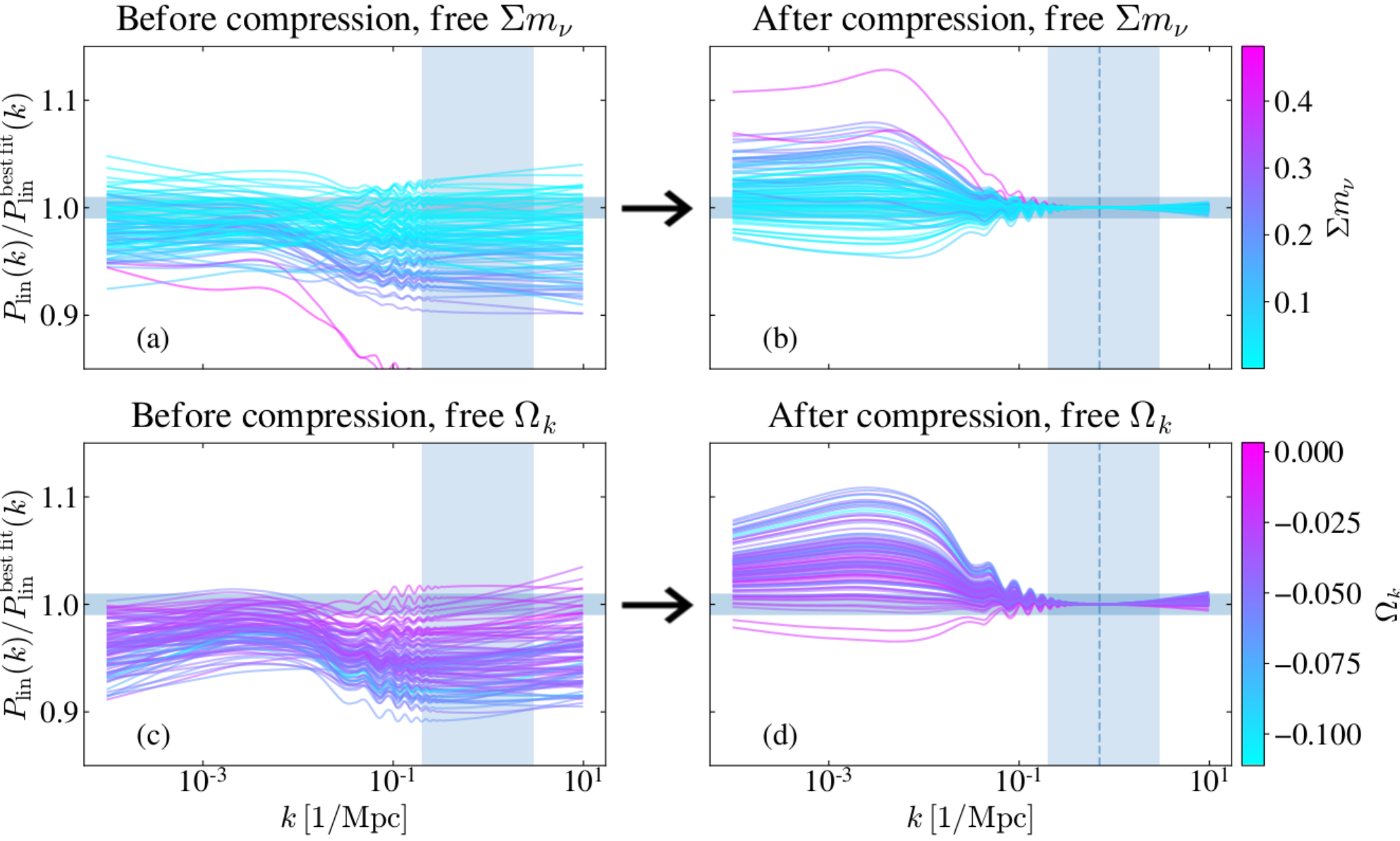}
    \caption{Information compression applied to two
    extended-$\Lambda$CDM parameter spaces.
    For 100 randomly sampled points in the \textit{Planck} chain,
    we plot the ratio of the $z=3$ linear matter power spectrum for each model
    with the baseline $\Lambda$CDM best fit model \cite{Planck2018}.
    In panel (a), we sample from a chain with free $\Sigma m_\nu$.
    In panel (b), for each of these models we rescale the amplitude and slope
    of the linear matter power spectrum to match at
    $k_p=0.7\, \mathrm{Mpc}^{-1}$, shown in the vertical dashed line.
    The colour of each line represents the value of $\Sigma m_\nu$ in that model.
    In the bottom panels we repeat the procedure for a chain with free $\Omega_k$.
    Panel (c) shows the ratio for 100 randomly chosen samples,
    and in panel (d) we have rescaled the slope and amplitude to match
    at the vertical dashed line.
    The vertical shaded area represents the length scales to which the \pd\ is sensitive
    in BOSS data \cite{Chabanier2019b}, and the horizontal gray band
    represents the region of $1\%$ agreement.}
    \label{fig:lin_rat}
\end{figure}

Figure \ref{fig:lin_rat} demonstrates why we consider just two parameters
describing a slope and an amplitude to be sufficient to parameterise
the linear matter power spectrum within this regime.
In panel (a), we randomly sample 100 points
from the \textit{Planck} chains\footnote{\url{https://wiki.cosmos.esa.int/planck-legacy-archive/index.php/Cosmological_Parameters}.} \cite{Planck2018}
with free \sigmanu, and plot the ratio of the linear matter power
spectrum at $z=3$ with the baseline $\Lambda$CDM best fit model,
which we denote as $P_\mathrm{lin}^\mathrm{\,best\:fit}$.
The colour of each line indicates the neutrino mass value for each model.
We see the expected behaviour where cosmologies with a higher
neutrino mass tend to lead to a suppression of the power spectrum
on small scales at late times.
In panel (b),
we take the same 100 samples and rescale
the amplitude and slope of the power
spectrum in order to match the small-scale linear power in
$P_\mathrm{lin}^\mathrm{\,best\:fit}$ as parameterised by
equations \ref{eq:Delta2_p} and \ref{eq:n_p}.
The vertical shaded region shows the approximate
length scales to which the \pd\ is sensitive in
BOSS data \cite{Chabanier2019b}.
The upcoming DESI results will provide an
improved measurement of the \pd\ over the same
range of wavenumbers, so the arguments presented
in this paper will continue to apply.
The horizontal band
shows the region of $1\%$ agreement.
The effects on the linear matter power spectrum of variations in the
7 free parameters are captured by changes to the slope and
amplitude to within $1\%$ (and often much better)
when considering only the length scales
within the shaded region.
This is compared with the most recent
constraints on the amplitude of the small-scale linear power
spectrum from BOSS
which have an uncertainty of $6\%$ \cite{Chabanier2019}.
In the bottom panels (c) and (d) we repeat the
procedure for a chain with free curvature parameter $\Omega_k$ 
instead of \sigmanu. Again we see that the all the variations
in cosmology are degenerate with changing just the slope and
amplitude of the power spectrum on the scales of interest.

A natural extension to $\Delta^2_p$ and $n_p$ would be to include a
parameter describing the second order derivative in the linear power around
$k_p$ (a running of the small-scale linear power spectrum).
The necessity of this parameter would be determined by the variation in the
running within the prior range of the cosmological models under investigation,
and the $k$ range and precision of the observational dataset.
It would also be possible to include an extra parameter
describing the growth rate if deemed necessary.
We leave such extensions to future work.

\subsection{Intergalactic medium}
\label{ss:igm}
The neutral hydrogen density is determined both by the
density of hydrogen (dependent upon the clustering
as described in the previous section, as well as the
mean baryon density $\omega_b$) and the ionisation state,
which is itself affected by several astrophysical factors.
A primary factor is the intensity of the ionising
metagalactic UV background (UVB), which is highly
uncertain. Assuming ionisation equilibrium this is
balanced with the rate of recombination, which is
dependent on the gas temperature. We parameterise
these effects in terms of an effective optical depth,
$\tau_\mathrm{eff}$, which is related to the mean
transmitted flux fraction $\langle F \rangle$ by
$\tau_\mathrm{eff}=-\mathrm{ln}\langle F\rangle$.

The thermal state of the gas around the mean cosmic density regions
that are probed by the \lyaf\ is well approximated by a power-law:
\begin{equation}
    T(\Delta_b)=T_0\Delta_b^{\gamma-1},
    \label{eq:tdr}
\end{equation}
where $\Delta_b=\rho_b/\bar{\rho}_b$ is the baryon overdensity , $T_0$ is the gas
temperature in $\mathrm{K}$ at mean density \cite{Lukic2015}
and \(\gamma\) is a free parameter.
The effect of $T_0$ on the recombination rate
is captured by $\langle F \rangle$, however thermal
motion of the gas particles also induces a
Doppler broadening of the absorption features.
This has the effect of smoothing the \pd\ on
small scales.
We describe this with a thermal broadening scale
\begin{equation}
    \sigma_T=9.1\sqrt{\frac{T_0}{10000}}\frac{1+z}{H(z)},
    \label{eq:sigt}
\end{equation}
where the factor $(1+z)/H(z)$ is required to convert the broadening
scale from $\mathrm{km/s}$ to $\mathrm{Mpc}$ (to match the units in
which we store the \pd).
We include the parameters $\sigma_T$ and $\gamma$ in order to
characterise the effects of gas temperature on the \pd.

On small scales the baryons are supported by gas pressure,
leading to a small-scale suppression in the baryon power
spectrum which therefore affects the \lyaf. 
Unlike the thermal broadening, this is a 3D effect
that also affects correlations perpendicular to the line of sight.
Following ref. \cite{Kulkarni2015} we parameterise this effect with
a pressure smoothing scale, $k_F$, in units of $\mathrm{Mpc}^{-1}$.

\section{Simulations}
\label{sec:sims}

To create a set of training data for constructing the emulator we present in section \ref{sec:emu}, we run a suite of 30 hydrodynamical simulations
using the \texttt{TreeSPH} code
\texttt{MP-Gadget}\footnote{\url{https://github.com/MP-Gadget/MP-Gadget}.} \cite{mp-gadget},
a variant of the widely used \texttt{Gadget-2} \cite{Gadget2} with improved
parallelisation and gravitational force solver.
The speed of these simulations is increased using the
\texttt{QuickLymanAlpha} option which
turns regions of baryon overdensity exceeding $10^3$ and with temperatures
$T < 10^5$ K into stars. These
highly non-linear regions are computationally expensive to evolve but do not
contribute to the \lyaf\ \cite{Viel2004b}.
Our simulation box size is $L=67.5$ Mpc with $768^3$ CDM and baryon particles.
These values were chosen as a compromise between modelling the low $k_\parallel$
modes that play an important role in cosmological constraints from BOSS data,
and modelling the small-scale effects of pressure smoothing of the IGM within
computational constraints.

\begin{figure}
  \centering
  \includegraphics[scale=0.31]{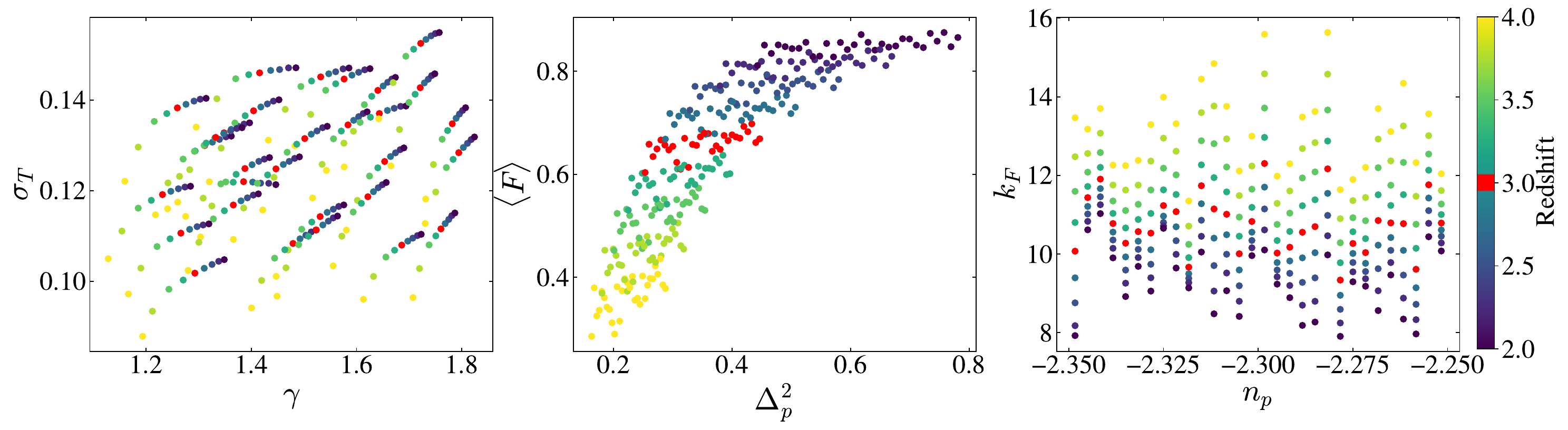}
  \caption{Distribution of the 270 training points in three example 2D projections.
  We colour training points by redshift purely for
  illustration, since we do not include redshift information
  in our emulator. The $z=3$ training points are highlighted
  in red for use in figure \ref{fig:red_test}.
  The Latin hypercube in simulation parameters shown in 
  table \ref{tab:sims} does not propagate into a Latin
  hypercube in emulator parameters, but ensures that the
  physically interesting parts of parameter space are
  populated with training points.
  }
  \label{fig:training}
\end{figure}

The simulations start at $z=99$ from initial conditions created
using \texttt{MP-GenIC}, which computes initial displacements using
the Zel'dovich approximation. The baryons and dark matter are initialised
on an offset grid
using species-specific transfer functions calculated in \texttt{CLASS} \cite{class}.
Following the simulations presented in ref. \cite{Pedersen2020},
in order to reduce the effects of cosmic variance we adopt the
`paired-and-fixed' approach from refs. \cite{Angulo2016,Anderson2018,FVN2018}.
In this method, the initial amplitudes of the Fourier modes are fixed,
with random phases.
The same random seed is used for all simulations, meaning that the residual
cosmic variance after applying the paired-fixed approach
is the same in each simulation.
We do not include any modelling of AGN feedback.
We output snapshots at $4>z>2$ at intervals of
$\Delta z=0.25$.
Note that this redshift information is not included in the emulator.
For each simulation snapshot we calculate the transmitted flux
fraction along $500^2$ lines of sight using the postprocessing
code
``\texttt{fake\_spectra}''\footnote{\url{https://github.com/sbird/fake_spectra}.} \cite{fakespec}.
We take the Fourier transform of the flux fraction along each line
of sight, and find the average of each Fourier mode to calculate the
\pd. The line-of-sight resolution is set to $0.05\,\mathrm{Mpc}$,
which is high enough to resolve the thermal broadening scale
$\sigma_T$ whose distribution is shown in figure \ref{fig:training}.
Box sizes and flux line-of-sight resolutions are chosen such that
our mock \pd\ have the same spacing in $k_\parallel$ for all
simulations and snapshots.

\begin{table}[]
  \centering
  \begin{tabular}{|l|l|l|l|l|}
  \hline
                    & Training simulations & \textit{central sim} & $h$ \textit{sim} & $\nu$ \textit{sim} \\ \hline
  $A_s(\times10^{-9})$  & $[1.35\mbox{--}2.71]$ &   2.00      &  2.01      &   2.25    \\
  $n_s$              & $[0.92\mbox{--}1.02]$ &  0.97 & 0.97 & 0.97   \\ 
  $h$             & 0.67 & 0.67 & 0.74 & 0.67    \\
  $\Omega_m$        & 0.316 & 0.316  &  0.259  &  0.324     \\ 
  $\Sigma m_\nu$  (eV)  & 0.0 & 0.0 & 0.0 &  0.3     \\ \hline
  $\Delta^2_p(z=3)$ & $[0.25$--$0.45]$ & \multicolumn{3}{c|}{0.35}   \\ 
  $n_p(z=3)$             & $-[2.35\mbox{--}2.25]$  &  \multicolumn{3}{c|}{$-2.30$}      \\ \hline
  $z_\mathrm{rei}$  & $[5.5\mbox{--}15]$ &  \multicolumn{3}{c|}{10.5} \\
  $H_A$  & $[0.5\mbox{--}1.5]$ &  \multicolumn{3}{c|}{1.0} \\
  $H_S$  & $[0.5\mbox{--}1.5]$ &  \multicolumn{3}{c|}{1.0} \\ \hline
  \end{tabular}
  \caption{Cosmological and astrophysical parameters for the training
  and test simulations. The limits of the Latin hypercube for the training
  simulations are shown in the left column, where only the primordial
  power spectrum and astrophysical parameters are varied.
  The primordial
  parameters $A_s$ and $n_s$ here are defined at the CMB pivot scale of
  $k=0.05\:\mathrm{Mpc}^{-1}$.
  The \textit{central sim},
  $h$ \textit{sim} and $\nu$ \textit{sim} test simulations are constructed
  such that they have the same small scale linear matter power spectrum
  ($\Delta^2_p$ and $n_p$) at $z=3$. For all simulations, we fix
  $\omega_c=0.12$, $\omega_b=0.022$.}
  \label{tab:sims}
\end{table}

In order to ensure that our training data are well distributed within the
range of physically interesting
models, the suite was set up using a Latin hypercube. The Latin hypercube spans the
parameters $[\Delta^2_p(z=3),n_p(z=3),z_\mathrm{rei},H_A,H_S]$, with
the ranges shown in the left column of table \ref{tab:sims}.
We use a flat $\Lambda$CDM cosmology, and vary only the primordial power spectrum
in order to sample $\Delta^2_p$ and $n_p$.
This will allow us to
test later our hypotheses from section \ref{sec:param}.
We fix $\omega_c=0.12$, $\omega_b=0.022$ throughout this paper, since they are
very well constrained by the CMB and otherwise poorly constrained by the \lyaf.
We also explore a range of IGM histories so that the astrophysical
parameters in section \ref{ss:igm} are well sampled. This is done
using modifications \cite{Onorbe2017} around a fiducial reionisation model presented
in ref. \cite{HM2012}, using a spatially uniform UVB.
We define $z_\mathrm{rei}$ as the midpoint of HI
reionisation, i.e. when the ionisation fraction of HI is 0.5, and
modify the UVB rates such that this point ranges between
$z_\mathrm{rei}\in[5.5,15]$.
Following ref. \cite{Bolton2008}, the thermal history is varied using
parameters $H_A$ and $H_S$, which respectively rescale the amplitude of the gas
heating rates, and the slope of their density dependence,
$\epsilon = H_A \, \epsilon_0 \, \Delta_b^{H_S}$.
These translate most directly into changes to $T_0$  (and therefore $\sigma_T$)
and $\gamma$ respectively.

In addition to the training simulations, we run three more
simulations that we use for various tests, shown in the right side of
table \ref{tab:sims}.
The first is a simulation
at the centre of the Latin hypercube of simulation parameters
in table \ref{tab:sims}.
We refer to this as the \textit{central sim}.
Second we run a $\Lambda$CDM simulation with a different background
expansion and growth, labelled \textit{$h$ sim}, with a larger value of
$h=0.74$ at fixed physical densities ($\omega_c$, $\omega_b$).
Finally we run a simulation
with $0.3\,\mathrm{eV}$ massive neutrinos, labelled \textit{$\nu$ sim}, where the neutrinos
are implemented using the linear response approximation
from ref. \cite{Haimoud2013}.
Again we keep $\omega_c$ and $\omega_b$ fixed.
The values of $A_s$ in these test simulations are all set such that
$\Delta^2_p(z=3)=0.35$, which is the midpoint of the Latin hypercube of
the training data. We use the same reionisation model in all test simulations.
These simulations are therefore constructed in such a way that they lie in
the same central part of emulator parameter space.
In the case of the \textit{$\nu$ sim}, this leads to an increase in $A_s$
to compensate for the suppression of the linear power on small scales caused by
massive neutrino free streaming \cite{Pedersen2020}.

The 30 training simulations, each with 9 snapshots, produce a total of 270 mock \pd\ measurements.
The Latin hypercube of simulation parameters in table \ref{tab:sims}
does not propagate into a Latin hypercube in the parameters
described in section \ref{sec:param}, but it does ensure
that the physically motivated regions of this parameter space
are well populated with training points.
In figure \ref{fig:training}
we show the distribution of these training points in three projections
of our six dimensional emulator parameter space. The points are coloured
corresponding to the redshift of each training point, although we show
this purely for illustrative purposes.
We assume that these 6 parameters are sufficient to describe the
\pd\ without including explicit redshift information,
and so we construct a single training set combining points from
all redshifts.
We highlight training points at $z=3$ in red for
discussion in section \ref{sec:emu}.

The trajectories of
individual simulations can be identified most clearly in the left
panel which shows the thermal parameters $\sigma_T$ and
$\gamma$. The middle panel shows points in the $\langle F \rangle-
\Delta^2_p$ plane. Both of these parameters are strongly
correlated with redshift, and so large parts of this
plane are unphysical, and therefore not populated with training points.
Although we do not explicitly include redshift information
in the emulator, we observe the redshift evolution of \pd\ through
the evolution of $\Delta^2_p$ and $\langle F \rangle$.
In the right panel we show the distribution of training
points in $k_F$ and $n_p$. There is no
redshift evolution of $n_p$, and so once again the
trajectories of individual simulations are identifiable.
The pressure smoothing scale evolves monotonically with redshift
as the gas accumulates thermal energy from the UV background.

\section{Gaussian process emulator}
\label{sec:emu}
The purpose of this paper is to model the \pd\ as a function
of the parameters described in section \ref{sec:param} using the simulations
presented in section \ref{sec:sims}.
To do this we use a Gaussian process (GP) \cite{gpml}
implemented in the Python package \texttt{GPy} \cite{gpy2014}.
Our simulations are constructed in such a way that every snapshot provides
a measurement of the \pd\ in 85 linearly spaced $k_\parallel$ bins covering
the range $[0.0931<k_\parallel<7.91]\:\mathrm{Mpc}^{-1}$.
Therefore we have a set of training points for each $k_\parallel$ bin,
and we drop the $k_\parallel$ notation for the following discussion.
We denote a point in parameter space as
$\bm{\theta}=[\Delta^2_p,n_p,\langle F\rangle,\sigma_T,\gamma,k_F]$.
A GP models the function one wants to learn, in this case
$P_\mathrm{1D}(\bm{\theta})$,
as a collection of random variables that form a joint Gaussian distribution, i.e.
$P_\mathrm{1D}(\bm{\theta})\sim\mathcal{N}(0,K(\bm{\theta},\bm{\theta}_i))$. Here
$\bm{\theta}_i$ are the locations of the training points where we have 
modelled
the \pd\ using simulations, and $K(\bm{\theta},\bm{\theta}')$ is a covariance kernel
whose precise form will be discussed shortly.
In order to approximate the Gaussian process prior of zero mean, we normalise
the \pd\ by the median value of the training data.
At a test point $\bm{\theta}_\star$, the joint distribution of the
test data $P_\mathrm{1D}(\bm{\theta_\star})$ and
training data $P_\mathrm{1D}(\bm{\theta}_i)$ can be written as
\begin{align}
    \begin{bmatrix}
        P_\mathrm{1D}(\bm{\theta}_i) \\
        P_\mathrm{1D}(\bm{\theta_\star})
    \end{bmatrix} &\sim \mathcal{N}\bigg(0,\begin{bmatrix}
        K(\bm{\theta}_i,\bm{\theta}_i)+\sigma^2_n\mathbf{I} & K(\bm{\theta}_i,\bm{\theta_\star})\\
        K(\bm{\theta_\star},\bm{\theta}_i) & K(\bm{\theta_\star},\bm{\theta_\star})
         \end{bmatrix}\bigg)
\end{align}
where $\sigma_n^2$ is a hyperparameter representing Gaussian noise on the 
training data. 
The mean and variance of the posterior predictive distribution given the
training information at $\bm{\theta}_i$ are then taken as our estimates of the
value and interpolation uncertainty for the $P_\mathrm{1D}(\bm{\theta_\star})$:
\begin{equation}
    \mu=K(\bm{\theta_\star},\bm{\theta}_i)[K(\bm{\theta}_i,\bm{\theta}_i)+\sigma^2_n\mathbf{I}]^{-1}P_\mathrm{1D}(\bm{\theta}_i)
\end{equation}
\begin{equation}
    \label{eq:gperr}
    \sigma^2=K(\bm{\theta_\star},\bm{\theta_\star})-K(\bm{\theta_\star},\bm{\theta}_i)[K(\bm{\theta}_i,\bm{\theta}_i)+\sigma^2_n\mathbf{I}]^{-1}K(\bm{\theta}_i,\bm{\theta_\star}).
\end{equation}
There is significant freedom in the choice of covariance kernel, which represents a weak
prior on the behaviour of the underlying function one wants to model. We use a squared
exponential kernel, also known as a radial basis function
\begin{equation}
    \label{eq:kernel}
    K(\bm{\theta},\bm{\theta}')=\sigma_0^2\mathrm{exp}\bigg(-\sum_{j}\frac{(\theta_j-\theta _j')^{2}}{2l_j^2}\bigg),
\end{equation}

with an independent correlation length $l_i$ in each parameter direction. It follows that our GP has a total of 8 hyperparameters:
6 correlation lengths, $\sigma_0^2$ and $\sigma_n^2$.
The hyperparameters are optimised by maximising the marginal log likelihood
of the training data \cite{gpml}. This is done across all $k_\parallel$ bins
simultaneously, so we use the same set of hyperparameters for the
full bandpower description of the \pd.
To avoid numerical issues, all of our parameters $\bm{\theta}$ are rescaled to vary within
a unit volume before being passed to the emulator.

We briefly highlight the differences in this setup compared to previous
applications of GPs to emulation of
the \pd\ \cite{Bird2019,Rogers2019,Rogers2020a}.
These emulators trained an independent GP on each redshift bin,
whereas we combine training points from all redshifts in one
GP. These works also used a different covariance kernel,
combining the squared exponential in equation \ref{eq:kernel}
with a linear term, and used an isotropic correlation length.
The parameterisation used in these works was such that
the \pd\ varied by a similar order of magnitude within the prior volume in each
parameter direction, meaning that an isotropic kernel
performed well.
A major difference
in our setup is that the sensitivity of the \pd\ to each
parameter varies strongly.
This is reflected in the optimised correlation lengths, which
range from $0.47$ for $\langle F\rangle$ to $11.5$ for $n_p$.
Since we combine all training
points in a single emulator and we use larger
simulations with more $k_\parallel$ bins,
there is more information in the training
set with which to optimise a larger number of hyperparameters
in an anisotropic kernel.

\subsection{Emulator validation}
\label{ss:val}
\begin{figure}
    \centering
    \includegraphics[scale=0.45]{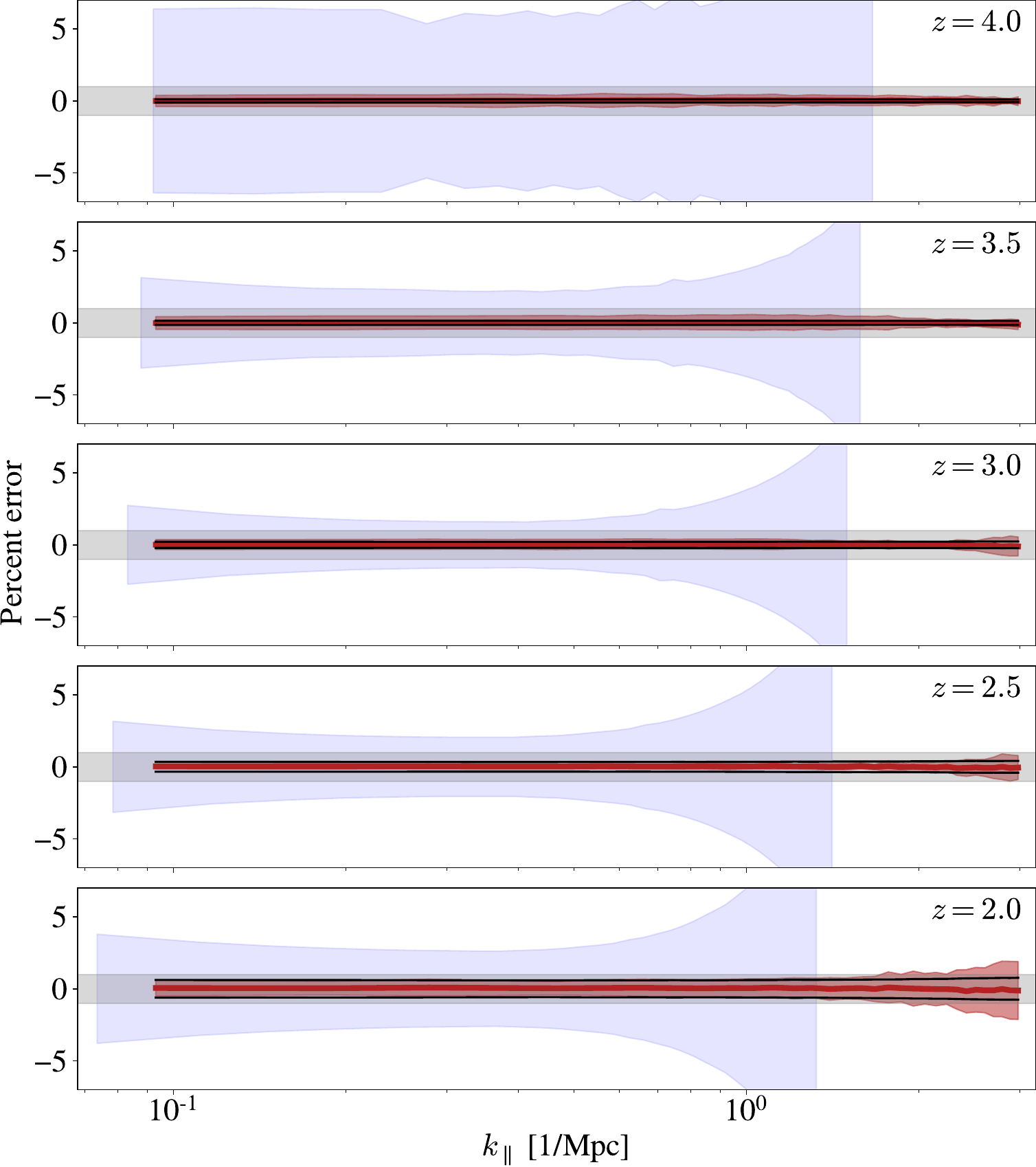}
    \caption{Results of a `leave-one-out' validation test on the GP emulator.
    For each of the 30 training simulations, we train the GP on the snapshots
    from 29 simulations and compare the GP prediction with the truth in
    each of the nine snapshots of the remaining simulation. 
    The solid red line and red shaded region respectively show the mean and standard deviation of
    the empirical error (prediction - truth) from the 30 validation tests.
    The black solid lines around zero show the $\pm$ median of the
    theoretical uncertainty on the emulator predictions
    (computed from equation \ref{eq:gperr}).
    The gray shaded area shows the region of $1\%$ error.
    We show the performance of the emulator at several redshifts to compare
    it with the observational uncertainties from the latest analysis of
    BOSS/eBOSS data (blue shaded regions \cite{Chabanier2019}).
    We expect that \pd\ measurements from DESI will cover a similar range
    of wavenumbers, but the systematic contribution to the data covariance is currently difficult to predict.
    }
    \label{fig:one_out}
\end{figure}

In this section we validate the emulator predictions on simulation data.
In figure \ref{fig:one_out} we show the results of a leave-one-out validation
test iterated across the full suite of 30 training simulations\footnote{This is in fact leave-\textit{nine}-out cross-validation, since we remove each time
from the training set all nine snapshots from a particular simulation.}.
We compare the \pd\ in the nine snapshots of each validation simulation with
the predicted \pd\ from an emulator trained on the rest of the suite.
The solid red lines and shaded red regions respectively show the mean and variance standard deviation of
the percentage error on these 30 predictions.
The black solid lines show the $\pm$ median $1-\sigma$ uncertainty on the
predictions as estimated by the GP using equation \ref{eq:gperr},
and the gray shaded area shows the region of $1\%$ error. 
We show the accuracy of the emulator predictions at different redshifts
to compare it with the uncertainty on BOSS measurements of the \pd\ from
ref. \cite{Chabanier2019}, represented by blue shaded regions\footnote{
For each panel we use the nearest redshift bin actually measured
in ref. \cite{Chabanier2019}.}.

The emulator predictions are unbiased, with a mean error of $<1\%$ at the 
redshifts and wavenumbers measured by BOSS, eBOSS and DESI.
The variance indicated by the shaded red region can be interpreted as an empirical
estimate of the average error on emulator predictions.
In all regimes probed by BOSS data this quantity is $<0.5\%$,
considerably lower than observational uncertainties.
The median theoretical error is slightly larger implying that
emulator uncertainties are slightly overestimated, which was
also the case in refs. \cite{Bird2019,Rogers2019,Rogers2020a}.
Nevertheless, these uncertainties are still also well below those of the BOSS data.
We do not include lines for DESI here as the exact increase in sensitivity
is still uncertain. However it is worth noting that in the case where DESI
errors become comparable to these uncertainties, there are several avenues
to be explored such as changes to the covariance kernel to improve emulator
performance or the addition of
refinement simulations using Bayesian optimisation \cite{Rogers2019,Rogers2020a}.

Besides the measurements from BOSS and eBOSS, the \lyaf\ \pd\ has also been
measured from small samples of very high-resolution spectra
\cite{Viel2013,Irsic2017,Yeche2017,Walther2018,Boera2019}.
These measurements extend to higher redshift and smaller scales, and e.g. can be
useful in breaking some of the parameter degeneracies (see appendix
\ref{app:param_depend}).
However, our simulations do not have the resolution required to resolve
the pressure scale for some of the colder IGM models at $z>4$
(see right panel of figure \ref{fig:training}).

\subsection{Predictions at a redshift without training data}
\label{ss:redtest}
In constructing a single GP using training points from all redshifts, we are assuming
that the \pd\ is sufficiently well described by the six parameters from section
\ref{sec:param}, without the need
to explicitly include redshift information in the emulator.
Under this assumption, if one were to construct
two simulations that had the same set of $\bm{\theta}$
parameters at different redshifts,
the \pd\ at these two different redshifts would be the same.
In practice, engineering such a simulation is non-trivial due to the complex relationship
between simulation parameters that model the IGM
and the astrophysical emulator parameters.
Therefore, in order to test
this hypothesis, we construct an emulator
dropping all training points at $z=3$ (the points
shown in red in figure \ref{fig:training}).
We then use this emulator to predict the \pd\ in the
$z=3$ snapshots for 3 randomly chosen
simulations, i.e. we are using training points at
$z\leq2.75$ and $3.25\leq z$ to predict the \pd\ at $z=3$.

\begin{figure}
    \centering
    \includegraphics[scale=0.4]{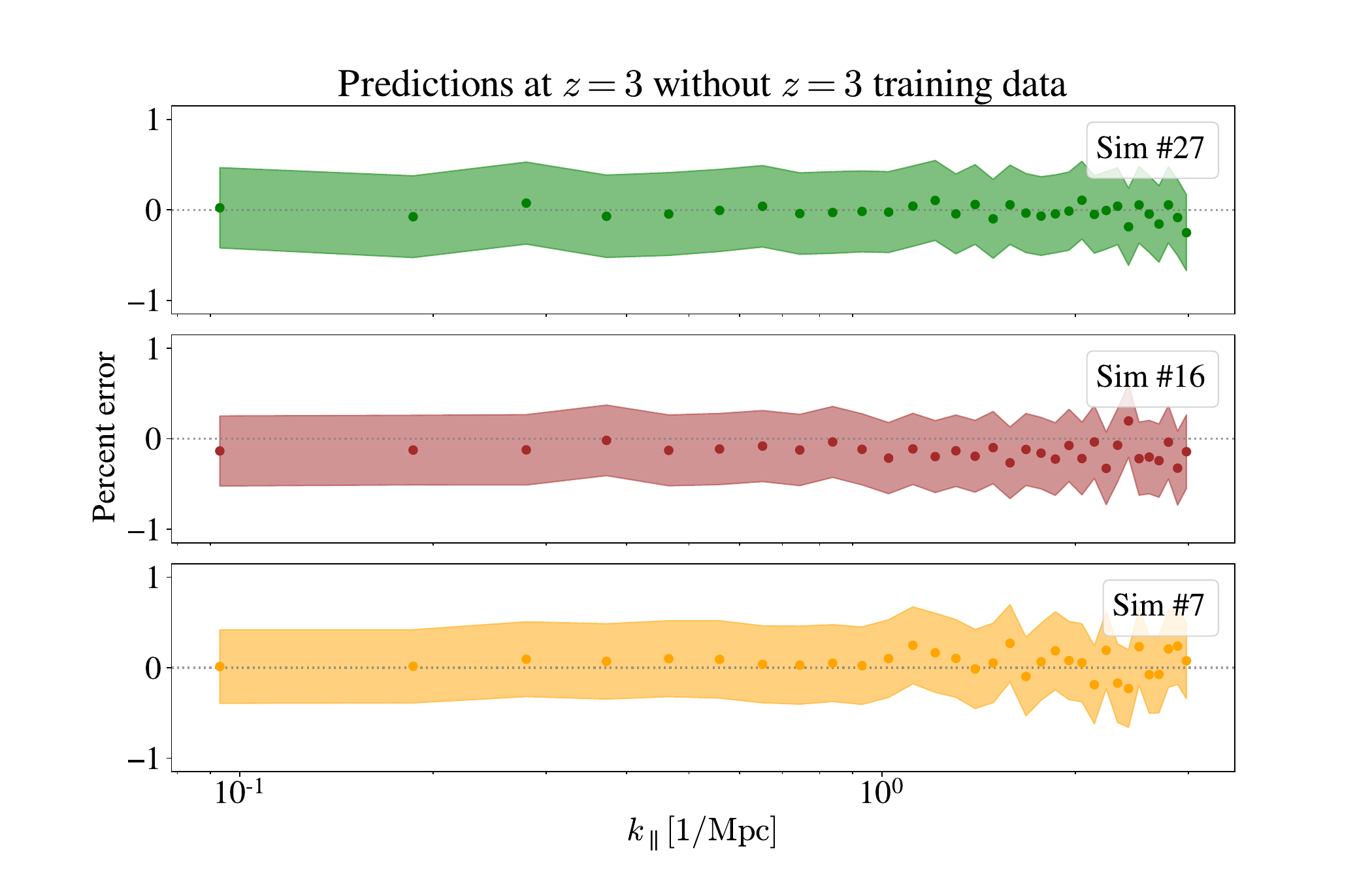}
    \caption{Emulator predictions for the \pd\ at $z=3$ when using
    no $z=3$ training points. We leave the red points from figure
    \ref{fig:training} out of the training set and reoptimise the
    GP hyperparameters. The three panels show the error on the
    emulated \pd\ for three randomly chosen simulations. The circles
    show the empirical error, and the coloured
    shaded region shows the $1-\sigma$ theoretical emulator uncertainty.}
    \label{fig:red_test}
\end{figure}

The accuracy of these predictions is shown in figure \ref{fig:red_test}, with the
theoretical uncertainty (computed using equation \ref{eq:gperr})
on each prediction shown in the shaded coloured region.
In each case
the predictions have sub-percent accuracy over the full $k_\parallel$ range.
This is a novel result that has not been possible with previous
\lyaf\ emulators, which validates our choice to combine training data
from all redshifts and omit redshift information.
The ability to emulate the \pd\ at arbitrary redshifts also confers
several advantages.
In particular, our emulator could be used to analyse DESI measurements of
\pd, regardless of the particular redshift binning of the data, which
is currently unknown.
Additionally, it is possible to combine training data from simulation
suites with different output redshifts, making the construction of training sets
more flexible.

\subsection{Predicting extended models}
\label{ss:extended}
Finally, we verify the emulator predictions on two test simulations: the \textit{$h$ sim}
and \textit{$\nu$ sim}.
The suite of training simulations has a fixed expansion history and
growth rate, with only the primordial power spectrum varied in order
to sample $\Delta^2_p$ and $n_p$.
By fixing the growth rate in all training simulations,
the emulator is learning the transfer function between the linear
matter power spectrum and the \pd\ for a fixed relationship
between the density power spectrum and the velocity power spectrum.
The velocity power spectrum will affect the \pd\ through redshift space
distortions (RSDs).
Therefore it is important to test the accuracy of the
emulator predictions in cosmologies with different
growth rates to the training set.
This will act as a test of our hypothesis that
the \pd\ is sensitive primarily to
the linear matter power spectrum and insensitive to minor changes in
the background expansion.

\begin{figure}
    \centering
    \includegraphics[scale=0.5]{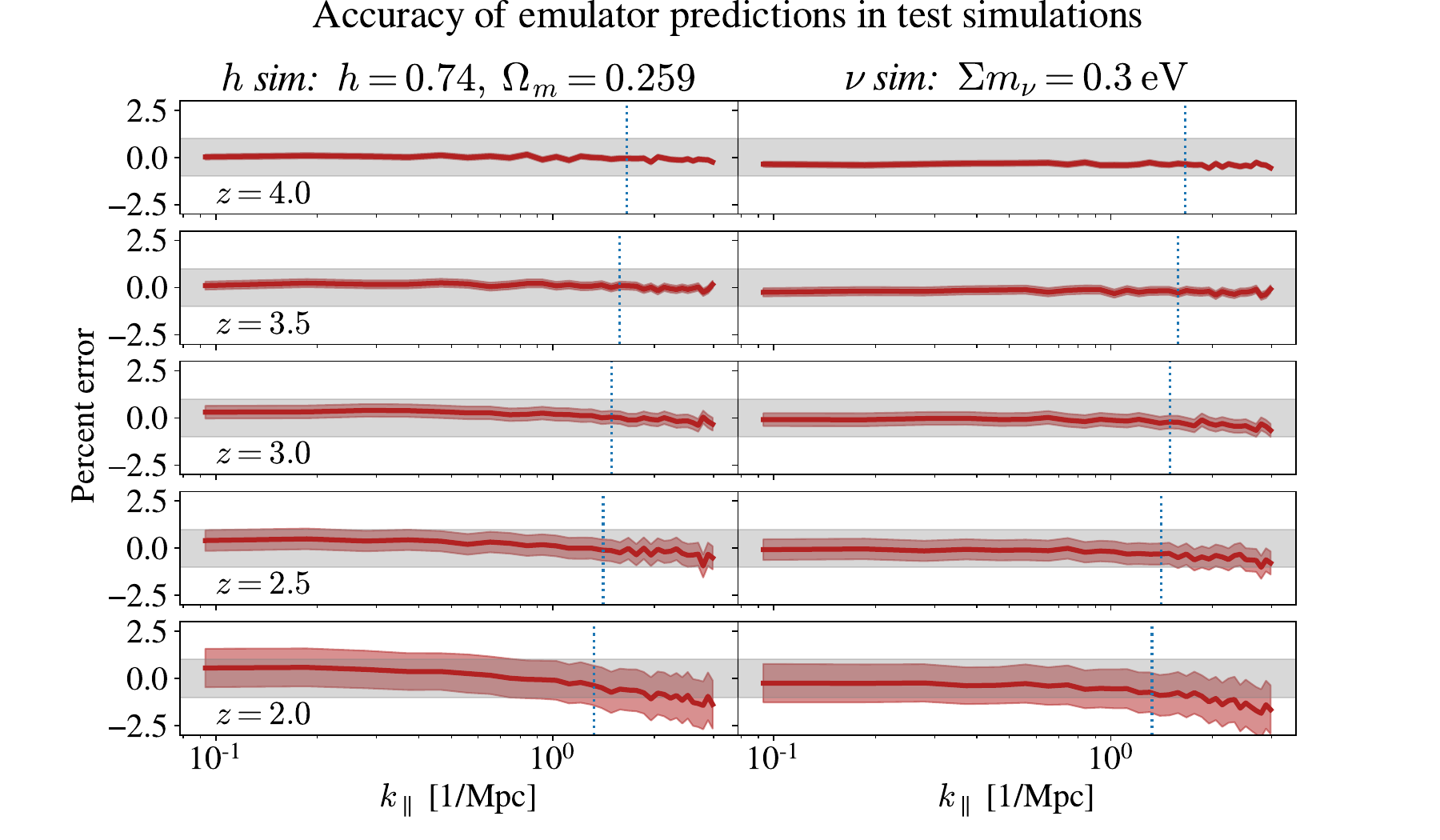}
    \caption{Accuracy of emulator predictions when tested on two test simulations from table \ref{tab:sims}.
    We test the emulator on different cosmological models to the training set, in which only
    the primordial power spectrum was varied.
    The red solid line shows the empirical accuracy of the emulator predictions, and the
    theoretical emulator uncertainty from equation \ref{eq:gperr} is shown in the red
    shaded region. 
    In the left panel we show predictions for the \textit{$h$ sim}, with $h=0.74$,
    where we have trained the emulator only on simulations with $h=0.67$.
    In the right panel, we show the accuracy of predictions for the \textit{$\nu$ sim}
    which includes $\Sigma m_\nu=0.3\:\mathrm{eV}$ massive neutrinos. The vertical dotted line
    shows the highest $k_\parallel$ bin of BOSS data, and the horizontal
    gray band shows the region of $1\%$ error.}
    \label{fig:nu_h}
\end{figure}

On the left side of figure \ref{fig:nu_h}, we show the accuracy
of the emulated \pd\ when compared with the truth in the
\textit{$h$ sim}.
The vertical dotted line shows the highest wavenumbers in BOSS
measurements \cite{Chabanier2019}, and the gray shaded area
shows the region of $1\%$ accuracy.
One way to consider this difference in $h$
within flat cosmologies
is as a variation in the onset of dark energy.
We are therefore
changing the growth factor in the redshift range
$4>z>2$, which will have two clear effects.
First, the redshift evolution of $\Delta^2_p$ will
change, and this effect is captured by our
parameterisation.
Second, the velocity power spectrum and therefore
the RSDs will be different to the simulations in the
training set. We are interested in determining whether
this effect, which is not captured by any of our parameters,
will impact the accuracy of the emulator predictions
to a statistically-significant degree.
Similarly to the tests shown in figure \ref{fig:one_out},
the predictions have sub-percent accuracy in the
regime probed by BOSS and DESI data.
Therefore we can conclude that the effect on the growth rate for this scale of
change in $h$ is negligible. 

In ref. \cite{Pedersen2020} it was shown that the effects of massive
neutrinos on the \pd\ are degenerate with changes to the amplitude of
the linear power spectrum to within $1\%$.
In that work it was difficult to disentangle effects of massive neutrinos on the
IGM, that were partially accounted for by changing $\langle F \rangle$ and
$\sigma_T$ in postprocessing. The emulator presented in this paper
allows for this degeneracy to be tested more clearly since the effects
of variations in the IGM and linear power spectrum can be predicted
independently.  The right side of figure \ref{fig:nu_h} shows the
accuracy of predictions compared with the truth in the \textit{$\nu$ sim},
which has $\Sigma m_\nu=0.3\:\mathrm{eV}$.
When adding massive neutrinos, we have kept $\omega_b$ and $\omega_c$ fixed.
Therefore we are modifying the late-time growth
and expansion rate governed by $\Omega_m$, which will have similar
effects to the previous case on the IGM, the growth of structure,
and baryon velocities.
Another effect is the suppression of late-time,
small-scale clustering due to free-streaming. This suppression is
fairly large - approximately $8\%$ in the case of
$\Sigma m_\nu=0.3\:\mathrm{eV}$, and it is this suppression that makes
the \lyaf\ a valuable probe of neutrino mass.  Again the effect on the
amplitude of clustering is captured in our setup by changes to
$\Delta^2_p$. 
We see similar results to the
previous test,
in that the emulator predictions are
accurate to less than 1 \%, with the best performance at
high $z$.
Note that this neutrino mass is larger than current bounds
$\Sigma m_\nu < 0.12$ ($95\%$ credibility)
from \textit{Planck} + baryon acoustic oscillations (BAO) \cite{Planck2018}.
In this extreme case the emulator still performs well; the
methodological errors for smaller neutrino masses
will be even smaller.
These results demonstrate that the relationship
between the linear matter power spectrum and the
\pd\ is insensitive to the specific background
expansion used in training the emulator.
The test simulations were set up such that
each emulator parameter is the same in both simulations
to within $2\%$.  This choice was made
so that the accuracy of
the predictions shown in figure \ref{fig:nu_h} serves
as a test of our parameterisation rather than as a test
of the interpolation.

\section{Parameter constraints}
\label{sec:sampler}
The ability of \lyaf\ information alone to constrain cosmological
parameters is limited by parameter degeneracies and insensitivities.
The constraining power of the \pd\ on cosmological parameters comes
when combined with CMB information, and we leave a discussion on this for future work.
However, in order to test the robustness of our emulator,
we present parameter constraints from a \lyaf-only likelihood on
the amplitude and slope of the primordial power spectrum,
$A_s$ and $n_s$ when using simulated mock data.
In order to minimise degeneracies, these quantities
are defined at a scale $k=0.7\:\mathrm{Mpc}^{-1}$.
To further test the assumptions of fixing the growth and expansion rate,
we also explore the sensitivity of the \pd\ to $H_0$.
Finally, we show constraints on $\Sigma m_\nu$ in order to investigate
the degeneracy between $\Sigma m_\nu$ and $A_s$ \cite{Viel2010,Pedersen2020}.
Given that constraints on $\Sigma m_\nu$ are a major motivation in \pd\ analysis,
understanding this degeneracy will be important in the analysis of DESI data.

\subsection{Likelihood parameters}
\label{ss:like_param}
Given that our emulator parameters are not associated with any redshift,
we do not attempt to constrain the emulator parameters themselves.
For our likelihood parameter vector $\thetal$ we consider two classes
of parameters -- cosmological and IGM parameters.
In section \ref{ss:samp_results} the cosmological parameters we vary are
$A_s$ and $n_s$, and we use uniform priors covering the ranges
$[1.1,2.5]\times10^{-9}$ and $[0.89,1.05]$ respectively.
In section \ref{ss:ext_models} we also vary \sigmanu\ and $H_0$, with uniform
priors covering the ranges
$[0,1]\,\mathrm{eV}$ and $[55,90]\:\mathrm{km}\:\mathrm{s}^{-1}\:\mathrm{Mpc}^{-1}$
respectively.
We keep the physical baryon and cold dark matter densities,
$\omega_c$ and $\omega_b$, fixed to the values
set in section \ref{sec:sims}.
For each combination of these parameters we use \texttt{CAMB} to calculate the
linear matter power spectra at each redshift where we have data, and we compute
the corresponding values of $\Delta^2_p(z)$ and $n_p(z)$.
Therefore each likelihood evaluation results in nine calls to the emulator.

To model the redshift evolution of the IGM parameters
we use the redshift evolution of each parameter in the
\textit{central sim} described in table \ref{tab:sims}
as a fiducial model. The evolution of the IGM for this simulation
and all 30 training simulations is shown in appendix \ref{app:igm}.
To vary the IGM, we use a multiplicative rescaling $m(z)$ 
of this fiducial
model using a power law around a pivot redshift:
\begin{equation}
    \mathrm{ln}\,m(z)=X_0+X_1\mathrm{ln}\,\frac{1+z}{1+z_\star},
    \label{eq:igm_model}
\end{equation}
where we set $z_\star=3$ as it is the midpoint of our redshift outputs.
$X_0$ controls the amplitude of the rescaling,
and $X_1$ accommodates changes in the redshift evolution of a given parameter.
The same approach is used for all IGM parameters, meaning we have a total
of eight free parameters to model variations in the IGM. This model will not
perfectly describe the evolution of the IGM in any simulation except the
\textit{central sim} and so we are not interested in constraints
on these parameters \textit{per se}, but instead interested in ensuring that we
allow for enough variation in the IGM for a thorough marginalisation.
The prior range for each parameter is $\in [-0.4,0.4]$, and we have
verified that marginalised constraints on cosmological parameters are
not impacted by expanding this range.

An advantage of omitting redshift information from the
emulator is the increased flexibility in modelling the redshift
evolution of the IGM. Since our emulator parameters are decoupled
from simulation and likelihood parameters, the redshift dependence
of the IGM is not fixed at the point of running the simulations
as is the case in many previous setups. Whilst we use
equation \ref{eq:igm_model} to vary the IGM in this analysis,
in principle any form of redshift evolution can be chosen provided
it does not involve calls to the emulator that lie outside the
convex hull of the training set.

Our baseline likelihood parameter vector is therefore
\begin{equation}
    \thetal=[A_s,n_s,\mathrm{ln}\,\tau_0,\mathrm{ln}\,\tau_1,\mathrm{ln}\,\sigma_0^T,\mathrm{ln}\,\sigma_1^T,\mathrm{ln}\,\gamma_0,\mathrm{ln}\,\gamma_1,\mathrm{ln}\,k_0^F,\mathrm{ln}\,k_1^F]
\end{equation}
where $\tau_i$ represents rescalings to the effective optical depth, $\tau_\mathrm{eff}$,
$\sigma^T_i$ to $\sigma_T$,
$\gamma_i$ to $\gamma$, $k^F_i$ to $k_F$.
The quantities that
relate to physical length scales, $\sigma_T$ and $k_F$
are evaluated in velocity units.

\subsection{Likelihood function}
\label{ss:like_func}
To more closely resemble a real analysis
we perform our likelihood evaluations in velocity units.
Since our emulator is constructed in comoving units,
a cosmological model is required to convert between these
using the relation $k=qH(z)/(1+z)$, where $k$ and $q$ are wavenumbers
in comoving and velocity units respectively, and denote a 
flux power spectrum in velocity units as
\pdv.
We use the same $q_\parallel$ bins as the BOSS data \cite{Chabanier2019}, and use a cubic
interpolation to rebin our mock and emulated flux power spectra to match
the BOSS bins. 
We use a Gaussian likelihood function:
\begin{equation}
    \label{eq:like}
    \mathrm{ln}\mathcal{L}(\thetal) = \sum_z\bigg[-\frac{1}{2}\Delta_i(q_i,z,\thetal)C_{ij}(z,\thetal)^{-1}\Delta_j(q_j,z,\thetal)-\frac{1}{2}\mathrm{log}\:\mathrm{det}\:C_{ij}(z,\thetal)\bigg]
\end{equation}
with
\begin{equation}
    \Delta_i(q_i,z,\thetal)=\tilde{P}_\mathrm{1D}^\mathrm{emu}(q_i,z,\thetal)-\tilde{P}_\mathrm{1D}^\mathrm{data}(q_i,z)
\end{equation}
where $\tilde{P}_\mathrm{1D}^\mathrm{emu}(q_i,z,\thetal)$ is the theoretical
prediction from the emulator for a given set of parameters $\thetal$,
and $z$ is the redshift at which we are making the calls to the emulator.
The covariance matrix $C_{ij}$ is a combination of the data
and emulator covariance:
\begin{equation}
    C_\mathrm{ij}(z,\thetal)=C_{ij}^\mathrm{data}(z)+C_{ij}^\mathrm{emu}(z,\thetal).
\end{equation}
In order to obtain results that resemble an analysis using current datasets,
for the data covariance $C_{ij}^\mathrm{data}(z)$ we use the
covariance matrices from the most
recent measurement of the \pdv\ from BOSS \cite{Chabanier2019}. Our
simulation snapshot outputs are different to the redshift binning
of the BOSS measurements, so for each $z$ where we have mock data,
we simply use the covariance matrix from the closest BOSS redshift bin,
prioritising the lower $z$ bin in the case where the bins are equidistant.
The emulator covariance matrix is constructed from the
theoretical uncertainty on each emulator call, which is a function of the
emulator hyperparameters and the local density of the training points.
Given that both these properties are common to all $q$ bins, we assume
the errors are maximally correlated and build
$C_{ij}^\mathrm{emu}(z,\thetal)$ under this assumption.
Since we use the same random seed in all simulations,
any residual cosmic variance that remains after applying the
`paired-and-fixed' approach is the same in all simulations.
We do not add noise realisations to the mock data drawn from the mock data
covariance.
The likelihood function is sampled using MCMC sampling implemented
in \texttt{emcee} \cite{emcee}, and we check for convergence of the
integrated autocorrelation time.

\subsection{MCMC results}
\label{ss:samp_results}

\begin{figure}
    \centering
    \includegraphics[scale=0.5]{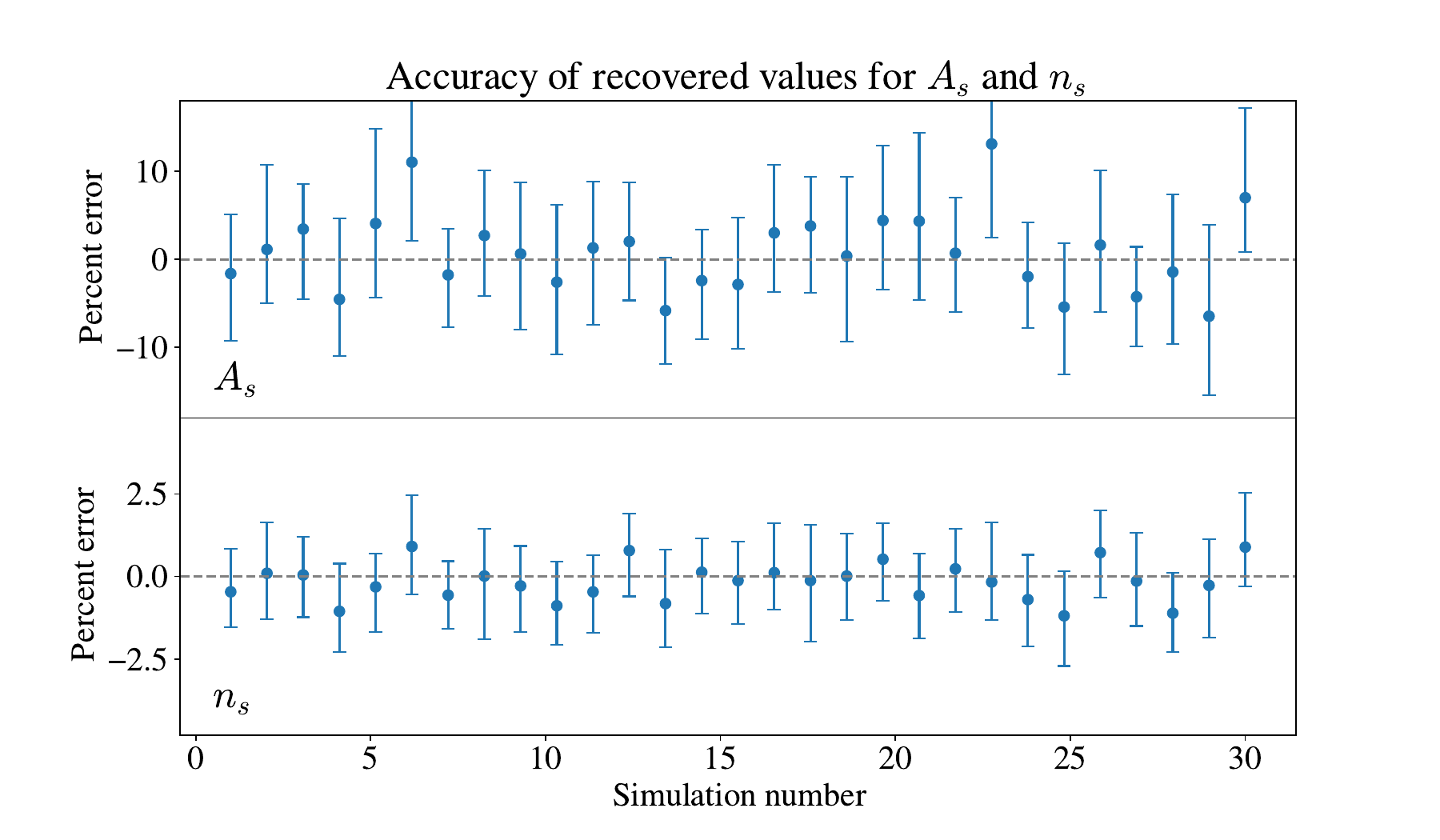}
    \caption{Accuracy of marginalised constraints on $A_s$ and $n_s$
    for each of the 30 training simulations, with $1-\sigma$ credible bounds.}
    \label{fig:fullsuite}
\end{figure}

In figure \ref{fig:fullsuite} we show the accuracy of marginalised
constraints on $A_s$ and $n_s$ for each of the 30 training simulations
when running with 10 free parameters:
$A_s$ and $n_s$ and the 8 IGM parameters described earlier. We fix
$h$ and \sigmanu\ to the true values of
$h=0.67$ and
$\summnu=0\,\mathrm{eV}$. As in section \ref{ss:val},
for each constraint we leave the validation simulation out of
the emulator training set. The error bars show the
$1-\sigma$ uncertainty. In all but 3 cases we recover the
correct cosmology to within $1-\sigma$.

\subsection{Extended models}
\label{ss:ext_models}

\begin{figure}
    \centering
    \includegraphics[scale=0.46]{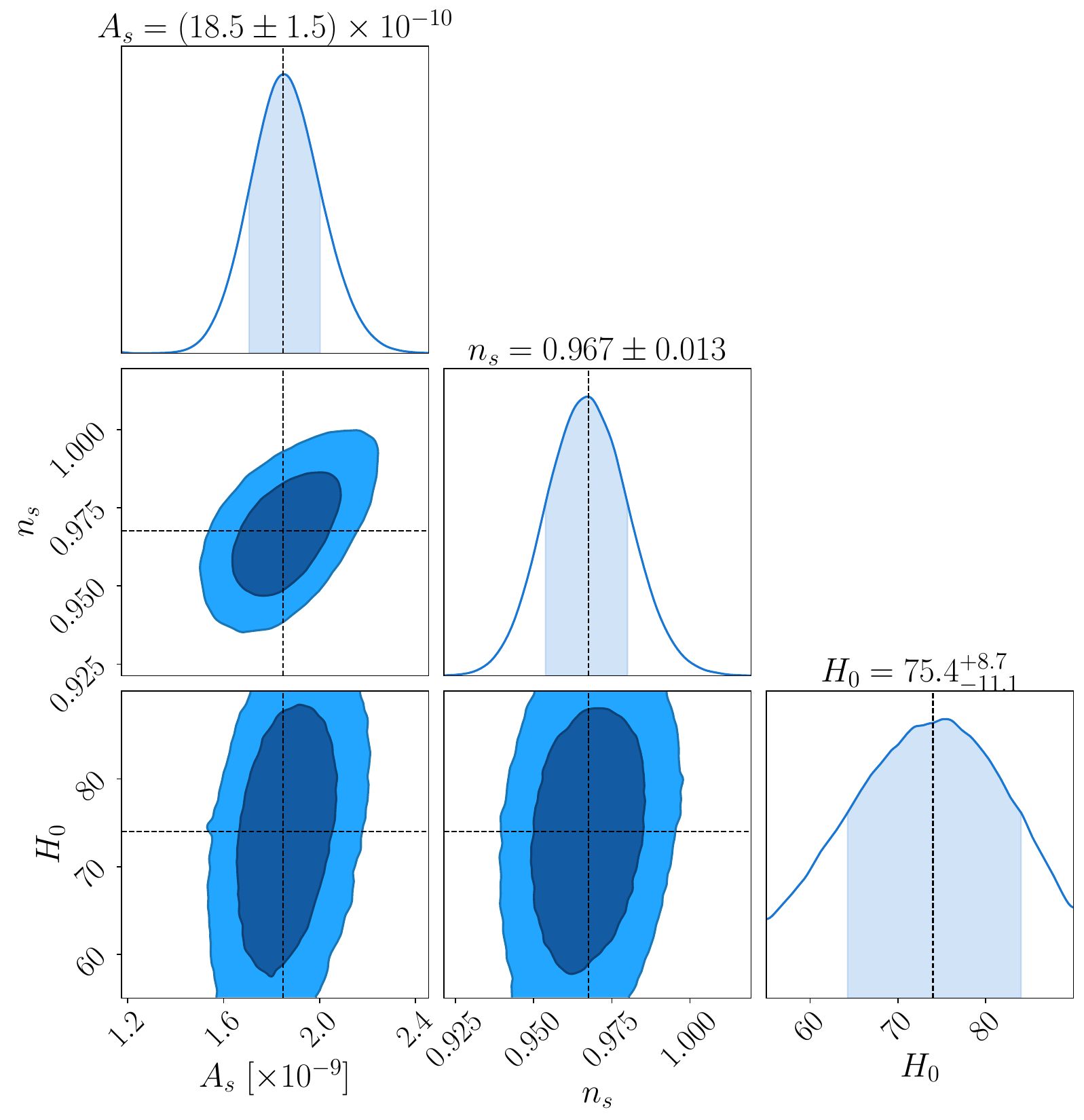}
    \caption{Marginalised 1D and 2D posterior distributions when using
    the \textit{$h$ sim} as mock data. The true
    values are indicated by the dashed lines, with
    the contours showing the $68\%$ and $95\%$ credible
    regions.
    $A_s$ and $n_s$ are defined at a pivot scale of $k=0.7\:\mathrm{Mpc}^{-1}$.
    The shaded area of the 1D posteriors shows
    the $68\%$ credible region.
    Posteriors for the remaining parameters are shown
    in appendix \ref{app:freeh}.
    }
    \label{fig:h_sampler}
\end{figure}

In figure \ref{fig:h_sampler} we show constraints
on $A_s$, $n_s$ and $H_0$ after marginalising over
the IGM parameters, using the \textit{$h$ sim} as mock data.
We set a wide prior range
of $H_0\in [55,90]\:\mathrm{km}\:\mathrm{s}^{-1}\:\mathrm{Mpc}^{-1}$.
Since the \lyaf\ is observed in velocity units,
our emulator predictions undergo a conversion
from comoving to velocity units using the
factor $H(z)/(1+z)$. Therefore we expect the
constraints on the amplitude of the power spectrum
to be degenerate with $H(z)$.
This is evident in the lower left
contour of figure \ref{fig:h_sampler}.
The degeneracy is weakened by the fact
that we are probing a high redshift where
the differences in $H(z)$ are smaller.
For example, across the wide prior
range, there is only a 
$6\%$ difference in $H(z=2)$, and
a $1.5\%$ difference at $H(z=4)$.
The constraints on $A_s$ and $n_s$ are
accurate. This is an important demonstration,
as we are correctly recovering parameters
in a cosmology
with a different background evolution to the one
used in constructing the emulator.

\begin{figure}
    \centering
    \includegraphics[scale=0.46]{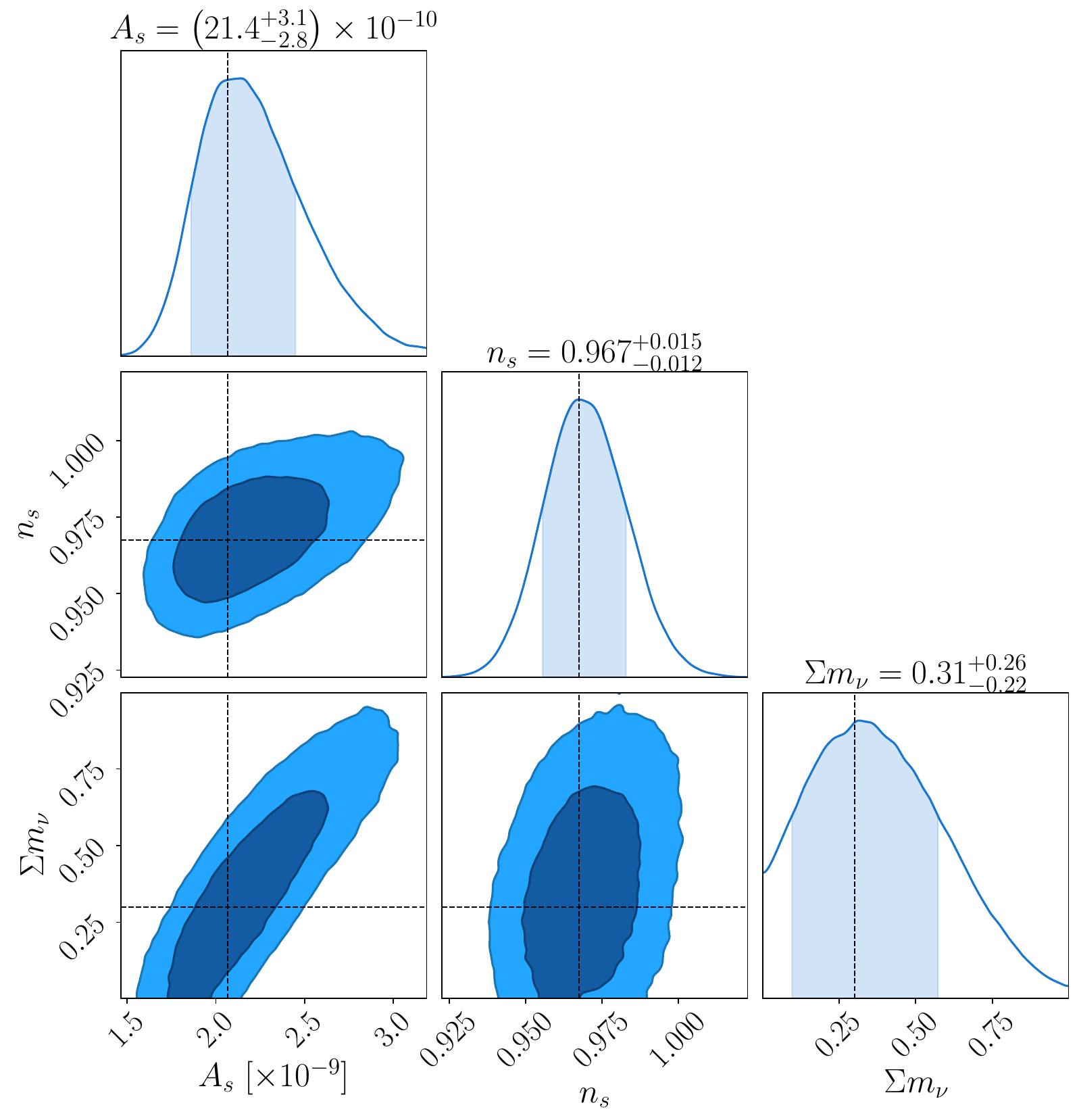}
    \caption{Marginalised 1D and 2D posterior distributions when using
    the \textit{$\nu$ sim} as
    mock data, with $\Sigma m_\nu$ as a free parameter. The true
    values are indicated by the dashed lines, with
    the contours showing the $68\%$ and $95\%$ credible
    regions.
    $A_s$ and $n_s$ are defined at a pivot scale of $k=0.7\:\mathrm{Mpc}^{-1}$.
    The shaded areas of the 1D posteriors show
    the $68\%$ credible region. For clarity in the figure, we do not show posteriors
    for the 8 free IGM parameters in this chain.}
    \label{fig:nu_sampler}
\end{figure}

In figure \ref{fig:nu_sampler} we show marginalised constraints
when using the \textit{$\nu$ sim}
as mock data, allowing neutrino mass to vary in the range
$\Sigma m_\nu \in [0.0,1.0]\,\mathrm{eV}$.
The effects of massive neutrinos
on cosmology are to vary both the expansion history and the growth
rate, and to suppress the amplitude of matter clustering on small
scales. Given the results of the previous figure, we expect the
effects of massive neutrinos on the expansion and growth to
have negligible contributions to the constraints.
We use an extreme value of $\Sigma m_\nu=0.3\,\mathrm{eV}$ that is
significantly higher than the current bound $\Sigma m_\nu<0.12\,\mathrm{eV}$ ($95\%$ credibility)
from joint \textit{Planck} + BAO data \cite{Planck2018}. 
The neutrino mass is strongly degenerate with $A_s$ when using the Lya forest alone \cite{Viel2010,Pedersen2020};
stronger bounds arise from a combination with CMB data by breaking the degeneracy (e.g \cite{Nathalie2015b,PD2020}).

\section{Discussion}
\label{sec:con}
We have presented a Gaussian process emulator that predicts
the \pd\ of the \lyaf\ as a function of the amplitude and slope
of the small-scale linear matter power spectrum and the
state of the intergalactic medium.
This work is an application of modern simulation interpolation techniques
to a modelling approach that was more prevalent in early analysis of the
\lyaf, based on the fact that the cosmological information in the \lyaf\ is
concentrated in the linear matter power spectrum at the observed redshift.
This is due to a unique property of the
\lyaf, in that it probes the universe at a period
very close to an Einstein de-Sitter model.
In this regime,
variations in the \pd\ due to differences
in the growth rate and expansion history (allowed by current data) are
negligible.
For the first time we test the accuracy of this approximation at
the level required by current datasets, and show that it holds.
In particular, we showed that even in models with massive neutrinos,
the cosmological information in \pd\ analysis
is contained in the slope and amplitude of the linear 
power spectrum within current observational uncertainty.

In addition to being lower dimensional in the cosmological
sector, our emulator affords other practical advantages.
By emulating the flux power spectrum independently of redshift, predictions
can be made to sub-percent accuracy at redshifts not necessarily included
in the training simulations.
This affords flexibility with regards to the analysis of future data.
Moreover, the redshift evolution of the IGM can be modelled at the point of
constructing a likelihood function, as opposed to being determined
at the point of constructing the emulator as in many previous frameworks.

Whilst we demonstrated that this emulator can be
applied to massive neutrino cosmologies without
massive neutrinos used in the training set, the same
principle can be applied to a wider range of extended cosmological models,
e.g. running of the spectral index in the primordial power spectrum
or curved universes \cite{Planck2018}.
There are two conditions under which our implementation
would not be valid. Models with sharp features in the
small-scale matter power spectrum (such as warm dark matter
models \cite{Viel2005,Viel2013} or ultra-light axion dark matter \cite{Hu2000,Hui2017}), or models with a non-standard
relationship between the baryon density and velocities (such as modified gravity
models \cite{Mustapha2019}).
However, simple extensions to the emulator parameter space
could incorporate some of these models too (e.g. \cite{Murgia2018,Rogers2020a}),
and the motivations for parameterising the \pd\ in terms of the
linear power spectrum still apply.

The joint analysis of CMB observations with measurements of the
small-scale linear matter power spectrum from the \lyaf\ has
long been a powerful combination in constraining extensions to $\Lambda$CDM
\cite{Phillips2001,Verde2003,Spergel2003,Seljak2005,Viel2004b,Seljak2006,Dvorkin2014,Krall2017,Archidiacono2019}.
We presented an emulator that will facilitate the combination of current (BOSS/eBOSS)
and upcoming DESI measurements of the \pd\ with CMB observations, to provide
competitive constraints on beyond-$\Lambda$CDM cosmologies.

\acknowledgments
The authors thank Jose O{\~n}orbe for sharing his code
for the reionisation model of ref. \cite{Onorbe2017} and for valuable discussions.
This work was partially enabled by funding from the UCL Cosmoparticle
Initiative.
This work used computing equipment funded by the Research
Capital Investment Fund (RCIF) provided by UK Research and
Innovation (UKRI), and partially funded by the
UCL Cosmoparticle Initiative.
This work was supported by collaborative visits funded by the Cosmology and
Astroparticle Student and Postdoc Exchange Network (CASPEN).
This work was performed using the Cambridge Service for Data Driven Discovery
(CSD3), part of which is operated by the University of Cambridge Research
Computing on behalf of the STFC DiRAC HPC Facility (\url{www.dirac.ac.uk}).
The DiRAC component of CSD3 was funded by BEIS capital funding via STFC capital
grants ST/P002307/1 and ST/R002452/1 and STFC operations grant ST/R00689X/1.
DiRAC is part of the National e-Infrastructure.
AFR acknowledges support by an STFC Ernest Rutherford Fellowship,
grant reference ST/N003853/1, and by FSE funds trough the program
Ramon y Cajal (RYC-2018-025210) of the Spanish Ministry of Science and Innovation.
AFR, HVP and AP were partially supported by the Science and Technology Facilities
Council (STFC) Consolidated Grant number ST/R000476/1.
KKR was supported by funding from the Science Research Council (VR) of Sweden;
and the Dunlap Institute for Astronomy \& Astrophysics,
University of Toronto.
PM was supported by the U.S. Department of Energy, Office of Science,
Office of High Energy Physics, under Contract no. DE-AC02-05CH11231.
HVP was additionally supported by the research project grant
“Fundamental Physics from Cosmological Surveys” funded by the
Swedish Research Council (VR) under Dnr 2017-04212.

\appendix
\section{Parameter dependence of the $P_\mathrm{1D}$}
\label{app:param_depend}
\begin{figure}
    \centering
    \includegraphics[scale=0.6]{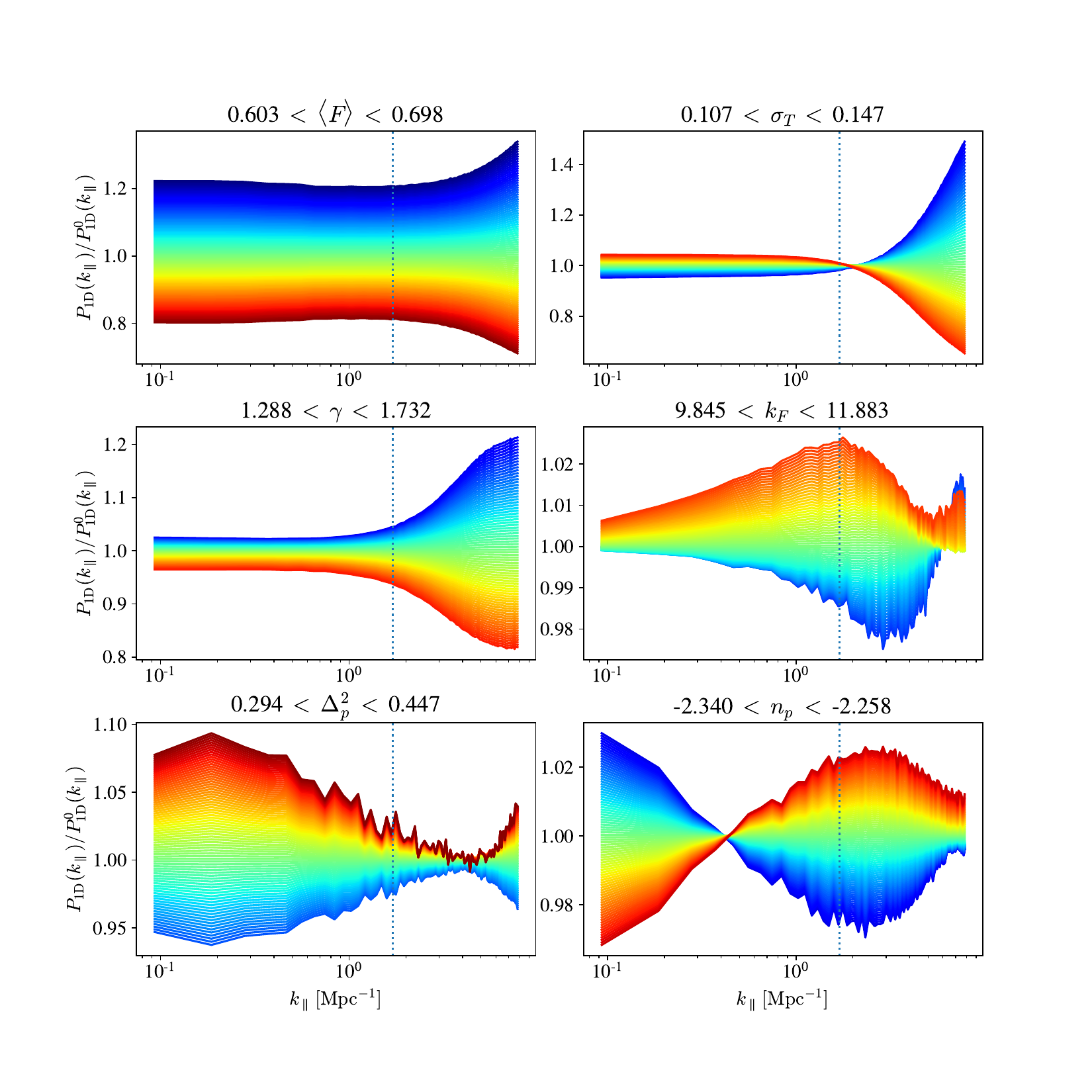}
    \caption{Fractional effect on the \pd\ of changing each of the six emulator
    parameters. We chose as a reference model, $P_\mathrm{1D}^0(k_\parallel)$
    the midpoint of the $z=3$ training points, and show the effect on the \pd\
    when moving from edge to edge of the convex hull of the training set.
    Lowest values for each parameter are coloured blue, transitioning
    to the highest values in red.
    The blue dotted line represents the highest $k_\parallel$ bin
    in BOSS data \cite{Chabanier2019}.
}
    \label{fig:param_depend}
\end{figure}

In figure \ref{fig:param_depend} we use the GP presented
in section \ref{sec:emu} to show the fractional change of the
\pd\ in response to changes of each of the parameters described
in section \ref{sec:param}. We show the fractional change around
a central point in parameter space set by the midpoint of all training
points at $z=3$, i.e. the points shown in red in figure \ref{fig:training}.
We denote this point as $P_\mathrm{1D}^0(k_\parallel)$.
We show the change in the \pd\ with respect to $P_\mathrm{1D}^0(k_\parallel)$
when moving from edge to edge
of the convex hull of the full training set, to ensure that we are not
extrapolating into regions of parameter space where the GP cannot reliably
predict the \pd.
The blue dotted lines show the highest $k_\parallel$
probed by BOSS \cite{Chabanier2019}.
The effects of IGM parameters are most dominant at higher wavenumbers
than those probed by BOSS data, with a degeneracy between
$\langle F \rangle$ and $\gamma$ when looking only at the
lower $k_\parallel$ modes.

\section{IGM evolution}
\label{app:igm}
\begin{figure}
    \centering
    \includegraphics[scale=0.6]{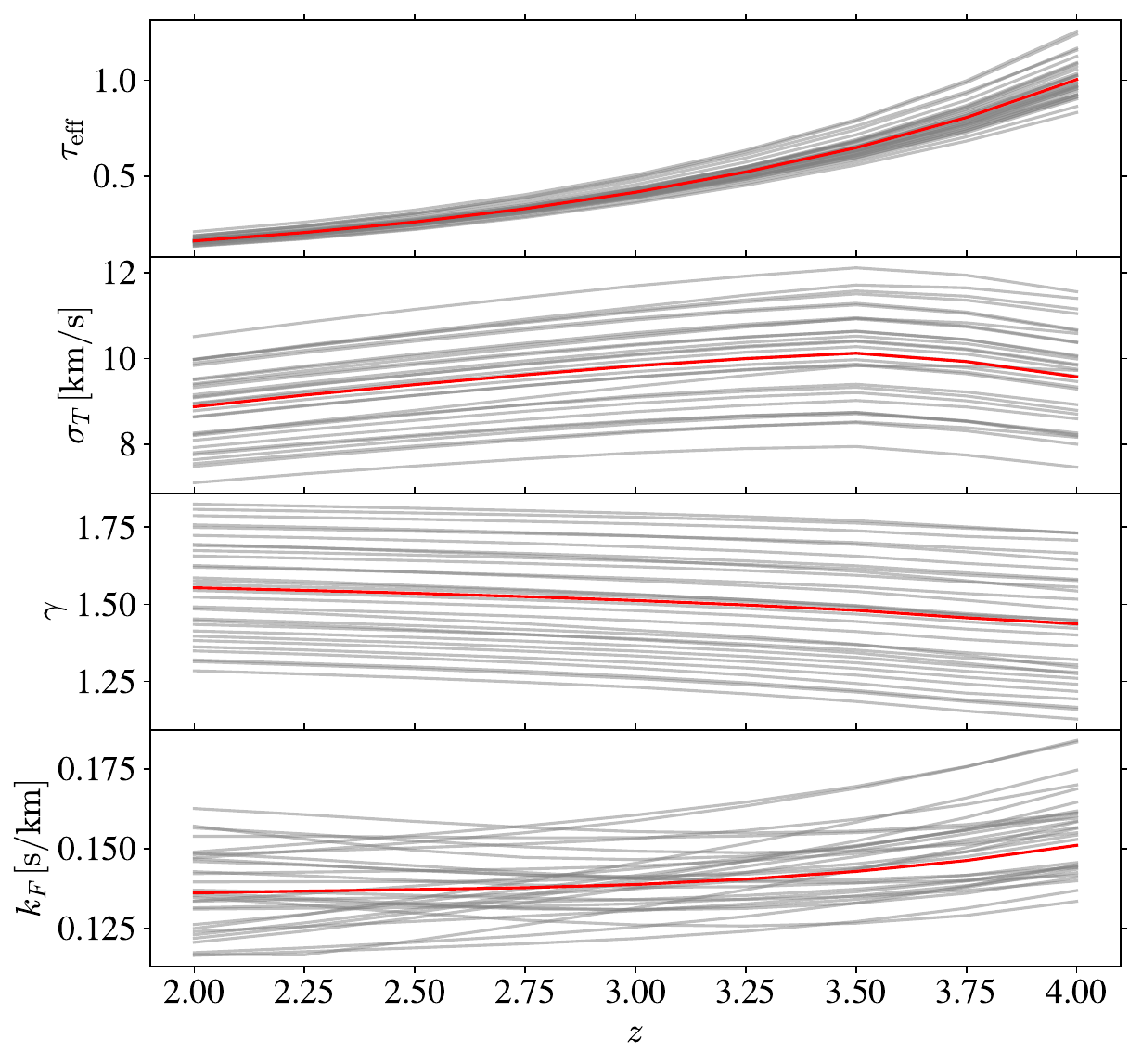}
    \caption{Redshift evolution of the IGM parameters for
    each of our 30 training simulations in gray. In red
    we show the IGM parameters for the \textit{central sim}
    which is used as a fiducial model in section
    \ref{sec:sampler}.}
    \label{fig:igmevol}
\end{figure}

We plot the redshift evolution of each IGM parameter for all 30
training simulations in gray lines in figure \ref{fig:igmevol},
with the central simulation shown in red.
The values in red are used to provide a fiducial model
for the marginalisation over the IGM described
in section \ref{sec:sampler}.
The mean flux, $\langle F \rangle$ is found by averaging
over the flux transmission fraction in every
line-of-sight pixel of a given snapshot.
To calculate the thermal parameters, $\sigma_T$
and $\gamma$, the particle snapshot data is binned in the
$\mathrm{log}_{10}T-\mathrm{log}_{10}\Delta_b$ plane, where $T$ is in units of $\mathrm{K}$.
We then perform a linear fit to the mode within the range $-1.5<\mathrm{log}_{10}\Delta_b<0.5$ to obtain values for
$T_0$ and $\gamma$ from equation \ref{eq:tdr}. $T_0$ is then converted into a value for $\sigma_T$ using
equation \ref{eq:sigt}.
In order to calculate the pressure smoothing scale, $k_F$, we follow
the approach in ref. \cite{Kulkarni2015}. We calculate the real-space
Lyman-$\alpha$ forest 3D
flux power spectrum, $\Delta^2_F(k)$, and fit an exponential cutoff
of the following form:
\begin{equation}
    \Delta^2_F(k)=Ak^n\mathrm{exp}\bigg(-\frac{k^2}{k^2_F}\bigg),
\end{equation}
where $A$ and $n$ are free parameters.

\section{Full posterior with IGM parameters}
\label{app:freeh}

\begin{figure}
    \centering
    \includegraphics[scale=0.4]{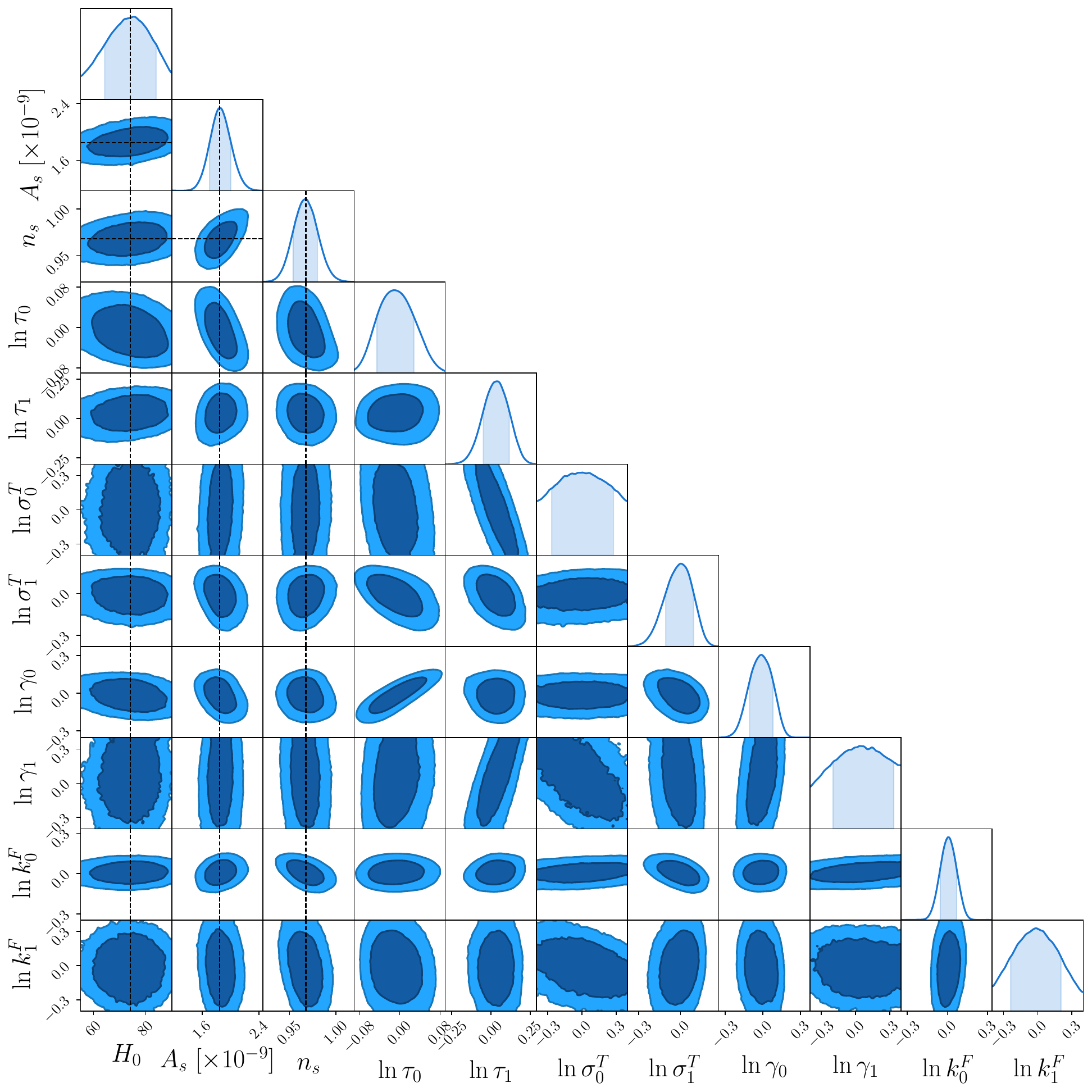}
    \caption{Marginalised 1D and 2D posterior distributions for all
    free parameters when using
    the \textit{$h$ sim} as mock data (the same as
    in figure \ref{fig:h_sampler}).
}
    \label{fig:fulligm}
\end{figure}

In section \ref{sec:sampler} we presented MCMC posteriors for several
analyses on mock data, after marginalising over the 8 IGM parameters.
In figure \ref{fig:fulligm} we show the full posterior for one of the analyses,
including cosmological and IGM parameters.

\bibliographystyle{JHEP.bst}
\bibliography{refs}
\end{document}